\documentclass[preprintnumbers,article,amsmath,amssymb,floatfix,10pt,prd,superscriptaddress,nofootinbib,twocolumn]{revtex4}

\usepackage{doi}
\usepackage{hyperref}
\hypersetup{
  colorlinks=true,        
  linkcolor=magenta,         
  citecolor=red,         
}

\usepackage{graphicx}
\usepackage{dcolumn}
\usepackage{bm}
\usepackage{color}
\usepackage{enumitem}
\usepackage{amsmath}
\usepackage{amssymb}

\newcommand{\beq}{\begin{equation}}
\newcommand{\eeq}{\end{equation}}
\newcommand{\bea}{\begin{eqnarray}}
\newcommand{\eea}{\end{eqnarray}}

\usepackage{bbm}
\usepackage{amsfonts}
\usepackage{mathrsfs}
\usepackage{latexsym}
\usepackage{epsfig}
\usepackage{epstopdf}
\usepackage{epstopdf}
\usepackage{graphicx}
\usepackage{amssymb}
\usepackage{amsmath}
\usepackage{dcolumn}
\usepackage{bm}
\usepackage{color}
\usepackage{comment}
\usepackage{xcolor}
\usepackage{subfigure}
\usepackage{orcidlink}

\begin{document}
\title{Light bending around the Kerr-Bertotti-Robinson black hole using material medium approach}

\author{Saswati Roy\, \orcidlink{0000-0002-7028-2627}}
\email{sr.phy2011@yahoo.com}
\thanks{Corresponding author}
\affiliation{Department of Physics, National Institute of Technology, Agartala, Tripura--799046, India}

\author{Anshul Tapase\, \orcidlink{0009-0000-7606-9942}}
\email{anshultapase@gmail.com}
\affiliation{Department of Physics, National Institute of Technology, Agartala, Tripura--799046, India}

\author{Shubham Kala\, \orcidlink{0000-0003-2379-0204}}
\email{shubhamkala871@gmail.com}
\affiliation{The Institute of Mathematical Sciences, C.I.T. Campus,
Taramani, Chennai--600113, Tamil Nadu, India}
\affiliation{Research Center of Astrophysics and Cosmology, Khazar University, Baku, AZ1096, 41 Mehseti Street, Azerbaijan}

\author{Hemwati Nandan\, \orcidlink{0000-0002-1183-4727}}
\email{hnandan@associates.iucaa.in}
\thanks{Corresponding author}
\affiliation{Department of Physics, Hemwati Nandan Bahuguna Garhwal University,
Srinagar--246174, Uttarakhand, India}
\affiliation{Centre for Space Research, North-West University,
Potchefstroom 2520, South Africa}

\author{Asoke Kumar Sen\, \orcidlink{0000-0003-2431-4525}}
\email{asokesen@yahoo.com}
\affiliation{Department of Physics, Assam University, Silchar, Assam 788011, India}

\begin{abstract}
In this paper, we study the deflection of massless particles due to a rotating, axially symmetric Kerr-Bertotti-Robinson (KBR) black hole via; material medium approach. We explored the effect of spacetime geometry on the trajectory of light rays in the presence of a uniform magnetic field. To derive an analytical expression for the deflection of light rays due to the Kerr-Bertotti-Robinson black hole, the frame dragging effect and refractive index were also studied in greater detail. From the analysis it is evident that the magnetic field actively adds to the black hole's gravity, making the bending of light stronger and permanently changing the space far away from the black hole, preventing it to act as a normal flat vacuum. From thermodynamical investigation, it is clear that entropy monotonically decreases with magnetic field strength and rotation parameter; whereas the Hawking temperature increases with a uniform magnetic field but decreases with spin parameter. 
\\

\end{abstract}

\maketitle

\date{\today}


\section{Introduction}

Black holes~(BHs) are one of the most fascinating solutions of General Relativity~(GR), which play a vital role in studies of the theory of gravitation, spacetime, and high-energy astrophysics~\cite{Einstein:1916vd,Schwarzschild:1916uq}. BHs, characterized by their event horizon from which no object can escape, are widely applied for testing of different gravitational theories in the regions of strongly gravitational fields~\cite{Weinberg72,Chandrasekhar:1985kt,Carroll:2004st}.
The recent revolutionary discoveries of gravitational waves and the direct observation of BH shadows helped us obtain a better knowledge of such compact objects and allowed us to test the theories of gravity in the strongly gravitational regime~\cite{LIGOScientific:2016aoc,Krolak:2017adi,EventHorizonTelescope:2019dse,EventHorizonTelescope:2022xqj}. These breakthroughs have marked a new age of gravitational physics observations, allowing researchers to probe the nature of compact bodies and certain long-standing problems related to strong-field gravity~\cite{Saumon:2022gtu,Luciano:2022eio,Branchesi:2023mws,Birrer:2025rdy}. As a result, BHs become excellent tools for probing the nature of spacetime, the behavior of matter in strongly gravitating environments, and even some extensions of GR.

One of the most astonishing effects predicted by the GR is the deflection of light rays in a gravitational field~\cite{Palatini:1923dza}. Specifically, the curvature of spacetime induced by a massive body results in the curvature of the light rays' trajectories causing the effect of gravitational lensing~\cite{Schneider:1992bmb}. The first experimental evidence of the light deflection in the gravitational field was obtained by Eddington during the solar eclipse expedition in 1919 when the shift in the position of some stars in the sky was measured and matched the prediction made by Einstein~\cite{Dyson:1920cwa}. Today, the effect of light deflection has turned into a valuable tool for studying the gravitational characteristics of compact objects. In this context, the most common tool for computing the light deflection is the null geodesic approach that assumes that the trajectories of photons (or any other massless particles) to be described by null geodesics defined by the underlying spacetime geometry~\cite{Synge:1934zza,Bardeen:1973tla}. Such an approach has been widely used for modeling the gravitational lensing of compact bodies like BHs and neutron stars~\cite{Fernando:2014rsa,Uniyal:2017yll,Alawadi:2020qdz,Vrba:2020mpt,Kala:2022uog,Kala:2025xnb}. Apart from the standard null geodesic approach, one may use the light deflection in a curved spacetime as an optical analog model. In such a scenario, the propagation of light waves is modeled as taking place within a medium whose refractive index depends on the gravitational field~\cite{Liberman:1992zz}. The impact of the gravitational field on the path taken by the photon is thus regarded as arising due to the changes in the refractive index that lead to bending of light in a way similar to refraction in normal optical media~\cite{tamm1924electrodynamics}. This optical model, popularly known as the \textit{Material Medium Approach}, offers another simple way of studying gravitational lensing effects and has been successfully used in many BH spacetimes.

The material medium model approach has become a very efficient alternative approach to studying the light propagation in a curved spacetime. For many years, it has found numerous applications in studying different gravitational effects such as the influence of rotating gravitating bodies on the state of polarization of electromagnetic waves, electromagnetic radiation scattering by gravitational fields, and optical characteristics of compact objects~\cite{Balazs:1958zz,Plebanski:1959ff,Mashhoon:1975ki}. Moreover, this model has been used to investigate the effect of gravitational lensing and light deflection in static, charged, and rotating BHs, shedding light on the effects of spacetime curvature via the concept of refractive index~\cite{sen2010more,Roy:2014hca,roy2015trajectory,roy2017deflection,roy2025deflection,roy2025non}. More recently, this approach has been expanded to a detailed analysis of Shapiro time delay and the interaction of gravitational waves with electromagnetic fields~\cite{delBarco:2025azz,Ruggiero:2024pqk}.

Astrophysical BHs are generally expected to exist in environments permeated by magnetic fields generated by surrounding plasma, accretion disks, and nearby magnetized matter~\cite{Fernandes:2025osu,Raha:2026pwt,Gupta:2026dzl}. Such magnetic fields are believed to play a crucial role in several high-energy astrophysical phenomena, including jet formation, particle acceleration, and the activity of active galactic nuclei~\cite{Abramowicz:2011xu,Davis:2020wea}. Consequently, the study of BHs immersed in external electromagnetic fields has attracted considerable attention in recent years. To describe these astrophysical scenarios, a variety of magnetized BH solutions have been constructed within GR, notably the Schwarzschild--Melvin, Kerr--Melvin, and Kerr--Newman--Melvin spacetimes, which provide exact models of BHs interacting with external magnetic fields~\cite{Hiscock:1980zf,Alekseev:1996fq,Booth:2015nwa}. The Bertotti--Robinson spacetime, an exact solution of the Einstein--Maxwell equations describing a homogeneous electromagnetic universe, also plays a fundamental role in BH physics owing to its close connection with the near-horizon geometry of extremal charged BHs~\cite{Dias:2003up,Cardoso:2004uz}. Motivated by these developments, the Kerr--Bertotti--Robinson (KBR) BH provides an interesting description of a rotating charged BH embedded in an external electromagnetic field~\cite{podolsky2025kerr}. This spacetime possesses several physically appealing features, including a well-behaved asymptotic structure and bounded ergoregions, making it a suitable framework for exploring the influence of external electromagnetic fields on particle and photon dynamics~\cite{Gibbons:2013yq,Bicak:2015lxa}. Although the solution appeared only recently, it has already stimulated extensive research activity. Several aspects of spacetime have been explored, including energy extraction through magnetic reconnection and the magnetic Penrose process, geodesic structure and shadow formation, optical characteristics, shadow-based parameter estimation, dynamics of spinning particles in external magnetic fields, the influence of surrounding matter distributions such as clouds of strings, accretion dynamics and quasi-periodic oscillations, as well as chaotic behavior in the strong-gravity regime~\cite{Zeng:2025olq,Wang:2025vsx,Zeng:2025tji,Ali:2025beh,Zhang:2025ole,Ahmed:2025ril,Mirkhaydarov:2026fyn,Singh:2026rbz}. The diversity of these studies highlights the rich physical properties of the KBR spacetime and its potential significance for both theoretical and observational BH physics.

The study of thermodynamical properties of BHs has been a significant aspect of gravitational theories ever since the seminal works of Bekenstein and Hawking established the formal correspondence between BH mechanics and classical thermodynamics~\cite{Hawking75}. BHs thermodynamics laws are very useful to understand the physical properties of a BH spacetime. In particular, they govern the intricate interplay between horizon area, Bekenstein–Hawking entropy, and Hawking temperature. These thermodynamic variables shows a dependence on the characteristic parameters of the BH, including the mass, angular momentum, charge and acceleration~\cite{Bekenstein73,Bardeen:1973tla}. Moreover, their behavior changes significantly due to external electromagnetic fields. Thus, this change reflects the influence of intrinsic BH geometry and the field distribution on the thermal and thermodynamic stability of BHs. More recently, Hu et al. investigated the thermodynamic properties of the KBR BH by employing the covariant phase-space formalism, establishing a consistent thermodynamic framework through the Christodoulou--Ruffini mass relation~\cite{Hu:2026slp}. Furthermore, Siahaan et al. explored the geometry and holographic aspects of the KBR BH and demonstrated the significant influence of the external Bertotti--Robinson electromagnetic field on its horizon and thermodynamic properties~\cite{Siahaan:2025ngu}. In this context, the thermodynamics of BHs in various alternative theories of gravity and BHs immersed in external fields has been extensively explored in the literature~\cite{Caldarelli:1999xj,Dehghani:2006ke,Yazadjiev:2013qna,Gibbons:2013dna,Anabalon:2018qfv,Rizwan:2023ivp,Shahzad:2024yrx,Bhattacharya:2024uce,Volovik:2025yzr,Babaei-Aghbolagh:2025qxm,ahmed2026charged}.

The propagation of massless particles in such spacetimes is especially significant because gravitational lensing provides a potential method for analyzing the geometrical features of compact objects. Apart from the standard null geodesic technique, light bending may also be analyzed through the material medium approach, where the gravitational field is described by an effective refractive index. Motivated by the unique properties of the KBR spacetime and the efficiency of the material medium approach, we study the gravitational bending of light in the KBR BH geometry. Furthermore, we examine the thermodynamic properties of the KBR BH to explore the impact of the external electromagnetic field on its thermal behavior and stability.

The organization of this paper is as follows. In Section~\ref{sec2}, we review the rotating KBR BH spacetime and discuss its main geometrical properties. In Section~\ref{sec3}, we derive the isotropic form of the KBR metric and obtain the corresponding refractive index, followed by an analysis of the frame-dragging effect. In Section~\ref{sec4}, we investigate the gravitational deflection of light in the KBR spacetime using the refractive index formalism. The thermodynamic properties of KBR BH are examined in Section~\ref{sec5}. Finally, in Section~\ref{sec6}, we summarize our main results and present the concluding discussion. Throughout this manuscript, we adopt geometrized units in which $G=c=1$ unless otherwise specified. However, in the analysis of the isotropic form and the frame-dragging effect, the speed of light $(c)$ is retained explicitly to preserve the physical interpretation of the corresponding expressions. For numerical computations and graphical illustrations, the BH mass is normalized to $m=1$.

\section{Rotating Kerr-Bertotti-Robinson Black Hole} \label{sec2}
The KBR solution describes a spacetime geometry that is entirely supported by a constant, uniform electromagnetic field. This exact solution to the Einstein–Maxwell equations exhibits remarkable symmetry and mathematical simplicity, making it one of the most elegant models for studying the interplay between gravitation and electromagnetism in GR. After its derivation, this solution has been considered in different contexts \cite{Zeng25,Kumar25,Ali26}. The line element describing the geometry of the rotating KBR BH is given by \cite{podolsky2025kerr,ovcharenko2025new}
\bea\label{BH}
ds^2 &=& \frac{1}{\chi} \left[(dt - \alpha \sin^2\theta d\phi)^2 \frac{Q}{\rho^2} - \frac{\rho^2}{Q}dr^2 - \frac{\rho^2}{P} d\theta^2\right. \nonumber \\ &-&\left. (\alpha dt - (\alpha^2 + r^2) d \phi)^2 \frac{P}{\rho^2} \sin^2\theta 
\right], 
\eea
where the metric functions take the form
\begin{eqnarray*}\label{MF}
\chi &=& \left(1+B^2 r^2\right)-B^2 \Delta \cos^2\theta,\\
Q &=& ( 1 + B^2 r^2 ) \Delta,\\
\Delta &=& \alpha^2+r^2 \left(1-\frac{B^2 I_2 m^2 }{I_1^2} \right)-\frac{2 I_{2} m r}{ I_{1} },\\
\rho^2 &=& r^2 + \alpha^2 \cos^2\theta,\\\
P &=& 1+ B^2 \cos^2\theta \left(\frac{I_2 m^2}{I_1^2}-\alpha^2\right),\\
I_1 &=& 1-\frac{\alpha^2 B^2}{2},\\
I_2 &=& 1 - \alpha^2 B^2.
\end{eqnarray*}

Here, $m$ is the mass of the BH (in the dimension of length), $\alpha$ is the rotation parameter of the BH, and $B$ represents the external uniform magnetic field parameter. For $B=0$, i.e. by vanishing the Maxwell field, the spacetime (\ref{BH}) reduces to the Kerr spacetime. Furthermore, setting the parameters $\alpha=0$ yields the result that the line element is reduced to the Schwarzschild metric and to the Bertotti Robinson Universe when $m=0$. In addition, $\alpha=0$ reduces the spacetime (\ref{BH}) to the Schwarzschild Bertotti Robinson (SBR) spacetime. 

The above spacetime has a spherical event horizon which can be determined with $\Delta= 0$ or $Q=0$, whose outer and inner horizons
are located at 
\begin{eqnarray} \label{horizon}
   r_\pm &=& \frac{mI_2 \pm \sqrt{m^2 I_2^2 - \alpha^2 I_1^2 + B^2 m^2 \alpha^2 I_2}}{I_1^2-B^2 m^2 I_2} I_1\nonumber\\
   &\approx&  \frac{mI_2 \pm \sqrt{m^2 I_2^2 - \alpha^2 I_1^2 }}{I_1^2-B^2 m^2 I_2} I_1.
\end{eqnarray}
From (\ref{horizon}) the horizons of Kerr BH can be recovered by setting $B=0$ and also for no rotation, i.e. $\alpha=0$ it becomes only one horizon as $r_{sch}=2m$. If we set only $\alpha=0$, we get $r_h=\frac{2m}{1-B^2 m^2} $, is the horizon of the Bertotti Robinson Universe which is greater than the horizon of the Schwarzschild BH.
The positive root ($r_H^+$) indicates the larger distance, which is the Event Horizon and the negative root ($r_H^-$) is the inner Cauchy Horizon.

\begin{figure}[h!]
    \centering
    \includegraphics[width=80mm,height=70mm]{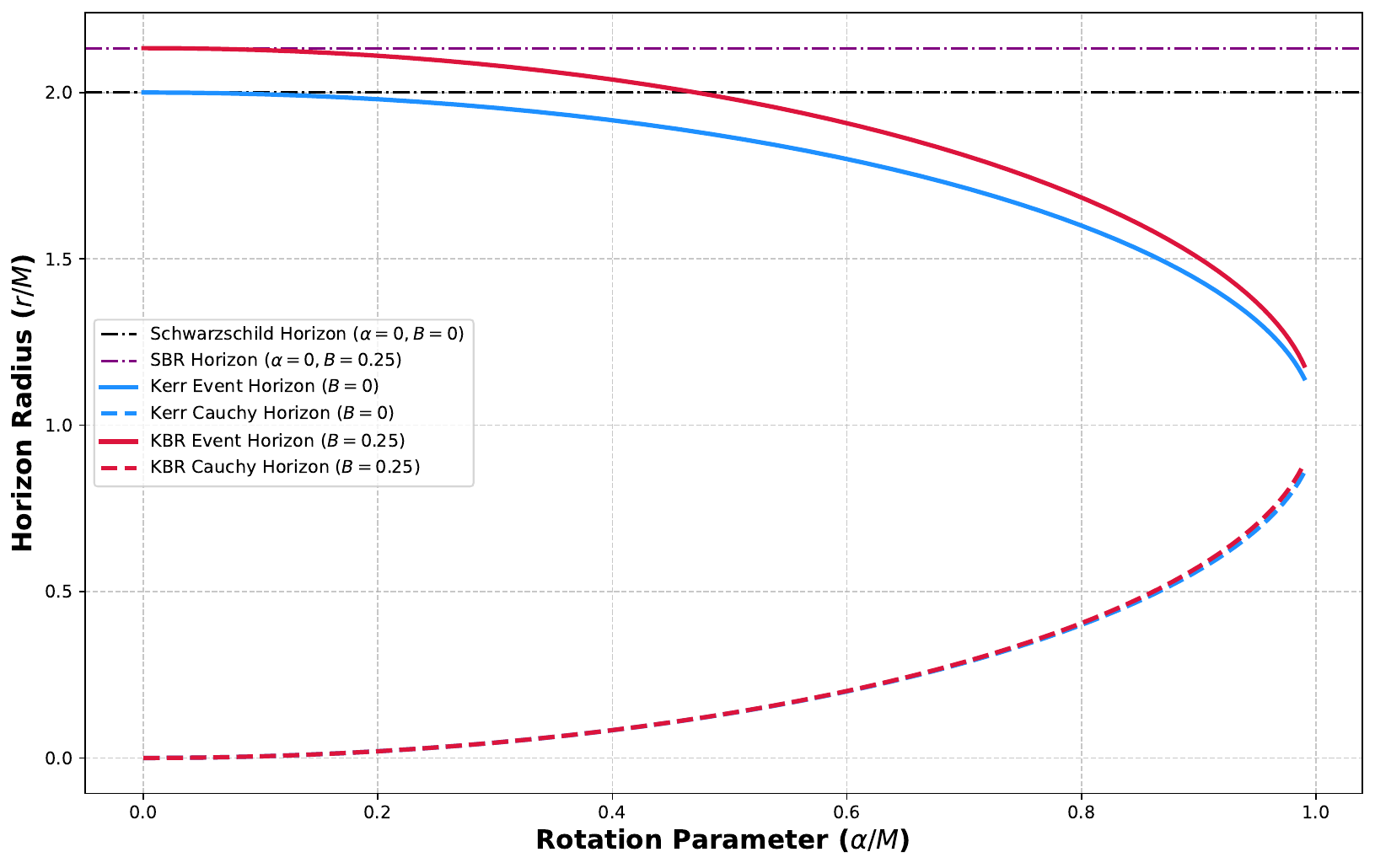}
    \caption{Variation of the Event Horizon ($r_H^+$) and the Cauchy Horizon ($r_H^-$) with respect to the rotation parameter $\alpha$ across unmagnetized ($B=0$) and magnetized ($B>0$) spacetimes .}
    \label{fig:Comparative_Horizons}
\end{figure}

In Fig. \ref{fig:Comparative_Horizons} we plotted horizons with the rotation parameter ($\alpha$) to clearly understand how rotation and the external magnetic field change the boundaries of the BH. 
This graph shows how both the outer event horizon ($r_H^+$) and the inner Cauchy horizon ($r_H^-$) behave with the change in the rotation parameter ($\alpha$). As the BH spins faster, the outer event horizon shrinks while the inner Cauchy horizon grows. When the rotation reaches its maximum limit, the two horizons merge into one. Adding a magnetic field clearly pushes the event horizon outward starting from the SBR limit. This acts as an expanded upper limit for the magnetized BHs.
An interesting geometric feature can be seen deep inside the black hole. While the magnetic field strongly pushed outward the outer event horizon, the inner Cauchy horizons ($r_H^-$) for both the Kerr and KBR BHs stay very close to each other, almost overlapping when the rotation is slow means that, right near the center, the extreme gravity and frame-dragging caused by the BHs mass and spin are so strong that the external magnetic field is completely overpowered.

\section{Isotropic form and Refractive Index due to Kerr-Bertotti-Robinson Black Hole} \label{sec3}
In the isotropic form of a metric, the spatial geometry is identical in all directions, with no distinction between the radial and angular components, apart from the coordinate dependence \cite{Weinberg72}. This means that no direction, whether radial or angular, is treated differently and distances scale uniformly regardless of which way you observe. In a general metric, the radial and angular parts can have different coefficients, which makes the geometry appear direction-dependent. The isotropic form avoids this by redefining the radial coordinate so that the spatial geometry becomes uniform in all directions. In an isotropic coordinate, as the spatial part is uniform in all directions, light behaves in a more symmetric way, which helps in understanding how gravity influences its path.

Rearranging and reinstating $G$ and $c$, the linearized form of the KBR metric (\ref{BH}) can be represented in terms of spherical polar coordinates ($ct, r, \theta, \phi$) as
{\small
\begin{equation}\label{BH1}
\begin{split}
ds^2 &= \frac{1}{\chi} \Bigg[
\left(
\frac{Q}{\rho^2}
- \frac{\alpha^2 P}{\rho^2}\sin^2\theta
+ \frac{2\alpha}{\rho^2}\big((r^2+\alpha^2)P-Q\big)
\sin^2\theta \frac{d\phi}{c\,dt}
\right)  \\
&\qquad \times c^2 dt^2 - \frac{\rho^2}{Q}dr^2
- \frac{\rho^2}{P}d\theta^2  \\
&\qquad - \left(
\frac{P}{\rho^2}(r^2+\alpha^2)^2\sin^2\theta
- \frac{Q\alpha^2}{\rho^2}\sin^4\theta
\right)d\phi^2
\Bigg].
\end{split}
\end{equation}
}
By letting $\theta=\pi/2$, we are reducing the metric down to the equatorial plane. Thus, in the equatorial plane, the line element (\ref{BH1}) becomes
{\small
\bea\label{BH2}
ds^2 &=& \frac{1}{\chi^{\prime}} \Bigg[
\left(
\frac{Q}{r^2} - \frac{\alpha^2}{r^2}
+ \frac{2\alpha}{r^2}\big((r^2+\alpha^2)-Q\big)\frac{d\phi}{c\,dt}
            \right)c^2 dt^2 \nonumber \\[4pt]
&-& \frac{r^2}{Q}dr^2 - r^2 d\theta^2 -\left(\frac{1}{r^2}(r^2+\alpha^2)^2 - \frac{Q\alpha^2}{r^2}
\right)d\phi^2
\Bigg],
\eea
}
where the metric functions changed as $\chi=\chi^{\prime}=(1+B^2 r^2)$, $\rho^2=r^2$ and $P=1$.

Now by applying the far field approximation (i.e. $\frac{\alpha^2}{r^2}<<1$) the KBR line element becomes 
\bea\label{BH3}
ds^2 &=& \frac{1}{\chi^{\prime}} \Bigg[
\left(
\frac{Q^{\prime}}{r^2}
+ 2\alpha \left(1-\frac{Q^{\prime}}{r^2}\right)\frac{d\phi}{c\,dt}
\right)c^2 dt^2
- \frac{r^2}{Q^{\prime}}dr^2 \nonumber \\[4pt]
&-& r^2\left(d\theta^2 + d\phi^2\right)
\Bigg],
\eea
with 
\begin{subequations}\label{eqn_Q_prime}
    \begin{align}
        Q^{\prime}(r)&=( 1 + B^2 r^2 ) \Delta^{\prime},\\
    \Delta^{\prime}(r)&=  \left(1-\frac{B^2 I_2 m^2 }{I_1^2} \right)r^2 -\frac{2 I_{2} m }{ I_{1} }r.
    \end{align}
\end{subequations}

To express the above line element (\ref{BH3}) in isotropic form, $r$ changes to $\mathcal{R}$ such that
\begin{equation} \label{Eq_drdR}
\begin{split}
    \frac{dr}{d\mathcal{R}}&=\frac{\sqrt{Q^{\prime}(r)}}{\mathcal{R}}.    
\end{split}
\end{equation}
As $Q^{\prime}$ is a function of $r$, introducing a new radial coordinate $\mathcal{R}$ cannot be obtained by any such simple artifice, but after a fair amount of tedious manipulation, it is found that
\begin{equation} \label{Eq_R}
    \mathcal{R}= \frac{1}{2}\left[r-\frac{b}{2a}+\sqrt{r^2-\frac{b}{a}r}\right]^{1/\sqrt{a}},
\end{equation}
with
\begin{eqnarray*}
    a&=&\left(1-\frac{B^2 I_2 m^2 }{I_1^2} \right),\\
    b&=& \frac{2 I_{2} m }{ I_{1} }.
\end{eqnarray*}
Considering $a=1$ (as the contribution of the second term of $a$ is negligible), the transformation equation (\ref{Eq_R})  becomes
\begin{equation} \label{Eq_R1}
    \mathcal{R}= \frac{1}{2}\left[r-\frac{b}{2a}+\sqrt{r^2-\frac{b}{a}r}\right],
\end{equation}
or,
\begin{equation}\label{Eq_r}
    r= \mathcal{R}(1+\frac{b}{4 \mathcal{R} a})^2.
\end{equation}
Now substituting the values of $r$ and $dr^2$ from Eqs. (\ref{Eq_r}) and (\ref{Eq_drdR}) in Eq. (\ref{BH3}) (which has a far-field or slow rotation approximation), we obtain the following isotropic form of the KBR line element.

\begin{eqnarray}\label{BH4}
ds^2 &= \frac{1}{\chi^{\prime}} \Bigg[
\left(
\frac{Q^{\prime}(\mathcal{R})}{\mathcal{R}^2(1+\frac{b}{4Ra})^4}
+ 2\alpha \left(1-\frac{Q^{\prime}(R)}{\mathcal{R}^2(1+\frac{b}{4\mathcal{R}a})^4}\right)\frac{d\phi}{c\,dt}
\right)c^2 dt^2
  \nonumber \\[4pt]
&- (1+\frac{b}{4\mathcal{\mathcal{R}}a})^4\left(d\mathcal{R}^2+ \mathcal{R}^2 \left(d\theta^2 + d\phi^2\right)\right)
\Bigg],
\end{eqnarray}

where 
\begin{eqnarray*}
    Q^{\prime}(\mathcal{R})&=&\left( 1 + B^2 \mathcal{R}^2(1+\frac{b}{4\mathcal{R}a})^4 \right) \Delta^{\prime}(\mathcal{R}),\\
    \Delta^{\prime}(\mathcal{R})&=& a \mathcal{R}^2(1+\frac{b}{4\mathcal{R}a})^4 -b\mathcal{R}(1+\frac{b}{4\mathcal{R}a})^2.
\end{eqnarray*}
In the KBR spacetime, rotation and a back ground electromagnetic field change the structure of spacetime in meaningful ways. As a result, the path of light is affected not just by the radial aspects of the metric, but also by the rotation of spacetime. In an  isotropic form, the spatial part becomes conformally flat. This setup provides a clear analogy to light travel in a medium. From this representation, we can identify the coordinate speed of light based on the metric’s null condition. Thus, from the isotropic form of the solution (\ref{BH4}), it is possible to obtain the refractive index of the effective medium using the material medium approach \cite{sen2010more, roy2015trajectory, roy2017deflection, roy2025deflection, roy2025non}. Unlike static spacetimes, the KBR metric includes off-diagonal terms related to rotation. These terms create a connection to the angular motion of the photon, resulting in a refractive index that is not just based on radial factors. Therefore, the effective refractive index reflects both gravitational and rotational effects on light travel. The refractive index expression relies on the metric functions of the KBR spacetime and implicitly on the angular speed of the photon’s path. This gives us a useful framework to examine how light bends and other optical effects occur when both rotation and electromagnetic fields are present.

Thus, the velocity of light $(v(r,\alpha, B))$ in the KBR spacetime can be identified as
\begin{eqnarray}\label{Eq_vel}
         v(r, \alpha, B)&=&c\frac{\sqrt{Q^{\prime}(r)}}{r}\sqrt{\frac{Q^\prime(r)}{r^2}+2\alpha\left(1-\frac{Q^\prime(r)}{r^2}\right)\frac{d\phi}{c dt}}\nonumber\\
         &=&c\frac{Q^{\prime}(r)}{r^2}\sqrt{1+2\alpha\left(\frac{r^2}{Q^\prime(r)}-1\right)\frac{d\phi}{c dt}}.
\end{eqnarray}
Under this formulation, the velocity of light is no longer a constant but becomes a function of position. This dependence arises due to the combined influence of rotation and the electromagnetic field present in the KBR spacetime. As a result, light propagation exhibits anisotropic behavior, which is essential to understand phenomena such as light bending.
\begin{figure}[h!]
    \centering
    \includegraphics[width=80mm,height=70mm]{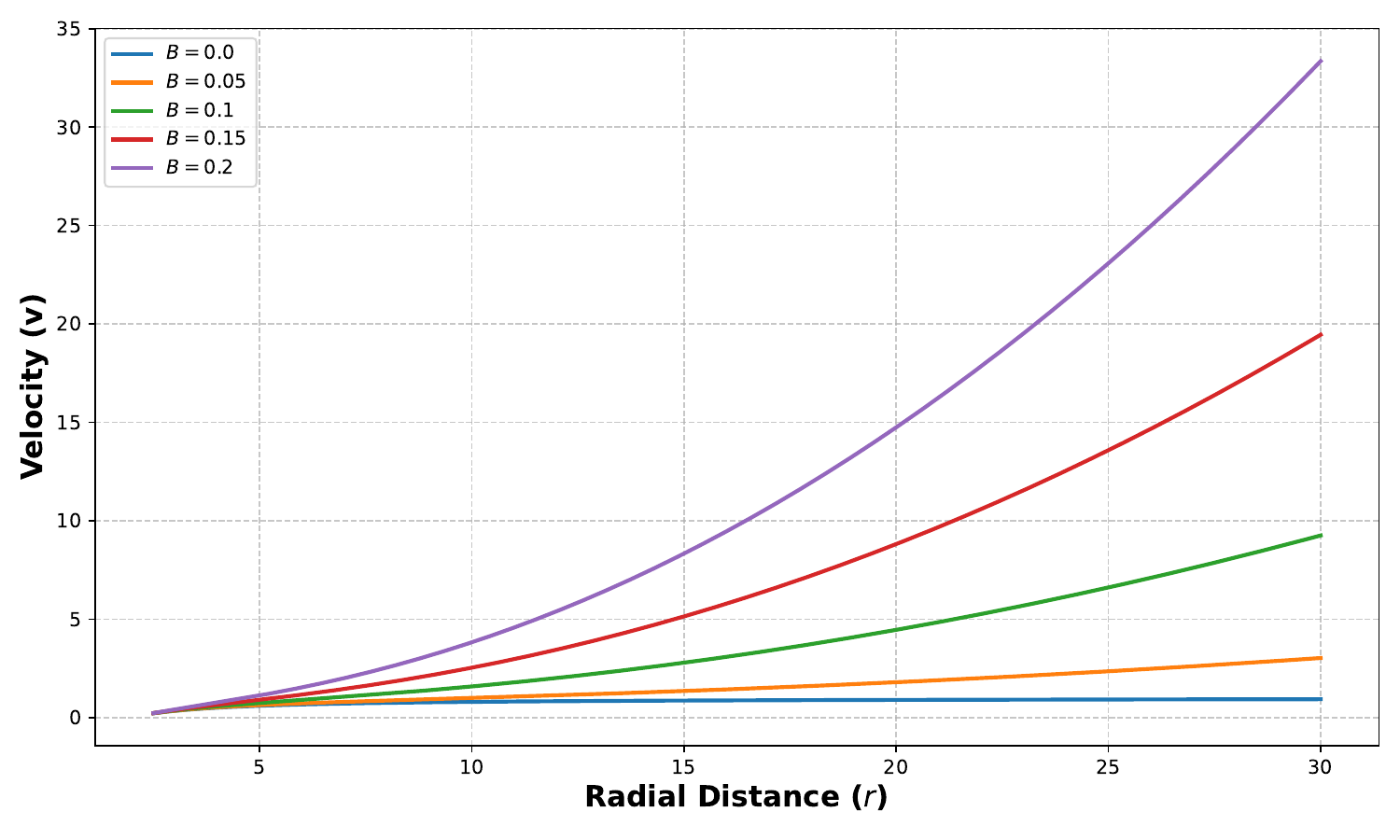}
    \caption{Effective velocity vs  radial distance graph for various values of parameter $B$. Here, we consider $\alpha = 0.5$.}
    \label{fig:velocity}
\end{figure}

The effective coordinate velocity graph ($v/c$) (fig. \ref{fig:velocity}) illustrates the profound optical impact of the KBR spacetime's magnetic background. For a standard Kerr BH ($B = 0$), the velocity asymptotically approaches $c$ at large distances, confirming that the spacetime reduces to a flat Minkowski vacuum \cite{Boyer67}. However, introducing a magnetic field ($B > 0$) causes the coordinate velocity to diverge quadratically. This proves the KBR metric is not asymptotically flat \cite{Wald10}; instead, it transitions into a Bertotti-Robinson universe where the omnipresent electromagnetic field structurally warps the far-field geometry.
Crucially, this apparent superluminal behavior does not violate General Relativity. The graph plots coordinate velocity ($\frac{dl}{dt}$), not local proper velocity, which always remains exactly $c$. Because the magnetic energy stretches the global spatial coordinate grid at large distances, light appears to cross artificially expanded coordinate distances in reduced coordinate time from the perspective of a distant observer. Within the Material Medium Approach, this demonstrates that the magnetic background acts as an atypical optical medium, fundamentally altering the effective refractive index and distinguishing the KBR metric's gravitational lensing signatures from those of a standard BH.

Now, the refractive index $n(r, \alpha, B)$ of the effective medium is
\begin{equation}\label{Eq_ref}
    n(r, \alpha, B)=\frac{r^2}{Q^{\prime}(r)}\left[1+2\alpha\left(\frac{r^2}{Q^\prime(r)}-1\right)\frac{d\phi}{c dt}\right]^{-1/2}.
\end{equation}
In the above expression (\ref{Eq_ref}), the refractive index depends not only on the radial coordinate through $Q^{\prime}(r)$ but also on the angular motion through the term $\frac{d\phi}{c dt}$, known as the frame-dragging parameter, which arises due to rotation of the gravitating mass. 
This reflects the influence of rotation in the KBR spacetime, leading to direction-dependent propagation of light.

In the absence of an external uniform magnetic field (i.e., at $B=0$), $Q^{\prime}(r)=\Delta^{\prime}(r)=r^2 - 2 m r $  and this expression corresponds to that of KBH \cite{roy2015trajectory}. Furthermore, if we set $\alpha=0$, we get the refractive index of the case of SBH \cite{sen2010more}. The contribution of the second term is significantly less than 1 for a slow rotating gravitating body; so the above expression (\ref{Eq_ref}) can be expanded into an infinite converging series.
\subsection{Frame-dragging}
When the central object is rotating, the spacetime is not just curved but also gets dragged along with the rotation \cite{Everitt11}. Because of this, particles and light moving
near the object are affected, even if they are not initially rotating. This effect is represented by the presence of the term $g_{t\phi}$ in the metric. As this term is not present in non-rotating spacetimes, so it directly indicates the role of rotation.

In a non-rotating case, time $t$ and angular coordinate $\phi$ are independent, meaning motion in time does not affect angular motion. However, in a rotating spacetime, the $g_{t\phi}$ term connects these two. Because of this, time
and angular motion are no longer separate, even if we try to keep $\phi$ constant, spacetime itself causes it to change, and the angular velocity $\frac{d\phi}{dt}$ changes 
depending on both energy and angular momentum together. 

The frame-dragging effect due to the Kerr–Bertotti-Robinson spacetime
can be obtained by using the four-momentum equation, as used in \cite{roy2015trajectory, roy2025deflection, roy2025non}. The four momentum of the particle can be written as ~\cite{Landau:1975pou}:
\begin{equation} \label{Eq_four_momentum}
     p^i=mc\frac{dx^i}{d\lambda}=g^{ik}p_k=-g^{ik}\frac{\partial S}{\partial x^k}.
\end{equation}
Here, i and k have values that go from 0 to 3, which stand for
the coordinates $ct$, $r$, $\theta$, $\phi$, respectively, and $\lambda$ is the affine parameter.
Also, the relativistic action function $S$, for a particle with time $t$ and angle $\phi$ as cyclic variables, in the gravitational field of a rotating spherical mass~\cite{Landau:1975pou}, is as follows
\begin{equation} \label{Eq_action}
    S=-E_0t+L\phi+S_r(r)+S_\theta(\theta).
\end{equation}
In the above equation $E_0$ denotes the conserved energy, while $L$ represents the part of the angular momentum along the field's symmetry axis and $S_r$, $S_\theta$ represent the parts of the action associated with the radius and the polar angle, respectively.

Compared with the Kerr–Newman line element expressed
by Eq.~(\ref{BH2}), the required co-variant components are
\begin{equation}
    g_{tt}=\frac{Q-\alpha^2}{\chi^{\prime} r^2},
\end{equation}
\begin{equation}
    g_{rr}=-\frac{r^2}{\chi^{\prime} Q},
\end{equation}
\begin{equation}
    g_{\theta \theta}=-\frac{r^2}{\chi^{\prime}},
\end{equation}
\begin{equation}
    g_{\phi \phi}=-\frac{(r^2+\alpha^2)^2 - Q\alpha^2 }{\chi^{\prime} r^2},
\end{equation}
\begin{equation}
    g_{t \phi}=g_{\phi t}=\frac{\alpha((r^2+\alpha^2)-Q)}{\chi^{\prime} r^2}.
\end{equation}
Correspondingly the contra-variant components are
\begin{equation}
        g^{tt}=\frac{\chi^{\prime} r^2((r^2+\alpha^2)^2-Q\alpha^2)}{(Q-\alpha^2)((r^2+\alpha^2)^2-Q\alpha^2)+\alpha^2(r^2+\alpha^2-Q)^2},
\end{equation}
\begin{equation}
    g_{rr}=-\frac{\chi^{\prime} Q}{r^2},
\end{equation}
\begin{equation}
    g_{\theta \theta}=-\frac{\chi^{\prime}}{r^2},
\end{equation}
\begin{equation}
    g^{\phi \phi}=-\frac{\chi^{\prime}r^2(Q-\alpha^2)}{(Q-\alpha^2)((r^2+\alpha^2)^2-Q\alpha^2)+\alpha^2(r^2+\alpha^2-Q)^2},
\end{equation}
{\small
\begin{equation}
    g^{t \phi}=g^{\phi t}=\frac{\chi^{\prime} \alpha r^2(r^2+\alpha^2-Q)}{(Q-\alpha^2)((r^2+\alpha^2)^2-Q\alpha^2)+\alpha^2(r^2+\alpha^2-Q)^2}.
\end{equation}
}
To obtain the value of frame dragging in the case of KSBH, we have generated the following equations from Eq. (\ref{Eq_four_momentum}) as
\begin{equation}\label{dphidt}
    \frac{d\phi}{c dt}= \frac{g^{\phi t}E_0-g^{\phi \phi}Lc}{g^{tt}E_0- g^{t \phi}L c}.
\end{equation}
\\
With the above equation (\ref{dphidt}) , the value of frame dragging ($\frac{d\phi}{cdt}$) is obtained as
\begin{eqnarray}\label{dphidt1}
    \frac{d\phi}{c dt}=&\frac{\alpha r^2(r^2+\alpha^2-Q) E_0 + r^2(Q-\alpha^2) Lc  }{r^2((r^2+\alpha^2)^2-Q\alpha^2) E_0 - \alpha r^2(r^2+\alpha^2-Q)  Lc}\nonumber\\
    =& \frac{\alpha r^2(r^2+\alpha^2-Q)  + r^2(Q-\alpha^2)\beta}{r^2((r^2+\alpha^2)^2-Q\alpha^2)  - \alpha r^2(r^2+\alpha^2-Q)  \beta},
\end{eqnarray}
\\
where $\beta$ is the impact parameter. As we have restricted the light ray in the equatorial plane, the angular momentum $L$ can be expressed as $L=p\beta$ and for light-like particles (i.e. photons), the momentum $(p)$ and the conserved energy $(E_0)$ are related as $E_0=pc$ which yields $\beta=\frac{Lc}{E_0}$. Now in the absence of an external uniform magnetic field (i.e. $B=0.0$), $I_1=I_2=0$ implies $Q(r) =Q^{\prime}(r)=\Delta^{\prime}(r)=r^2 - 2 m r $; and the frame dragging effect corresponds to that of KBH \cite{roy2015trajectory}.

Thus, the refractive index (\ref{Eq_ref}) with the frame dragging value, obtained as (\ref{dphidt1}), can be expressed as

\begin{eqnarray}\label{Eq_ref_final}
    n(r, \alpha, B)=&\frac{r^2}{Q^{\prime}(r)}\Bigg[1+2\alpha\left(\frac{r^2}{Q^\prime(r)}-1\right) \nonumber\\
    \qquad & \times \frac{\alpha r^2(r^2+\alpha^2-Q)  + r^2(Q-\alpha^2)\beta}{r^2((r^2+\alpha^2)^2-Q\alpha^2)  - \alpha r^2(r^2+\alpha^2-Q)  \beta}\Bigg]^{-1/2}.
\end{eqnarray}
\begin{figure}
    \centering
        \begin{subfigure}[]
           {\includegraphics[width=90mm,height=70mm]{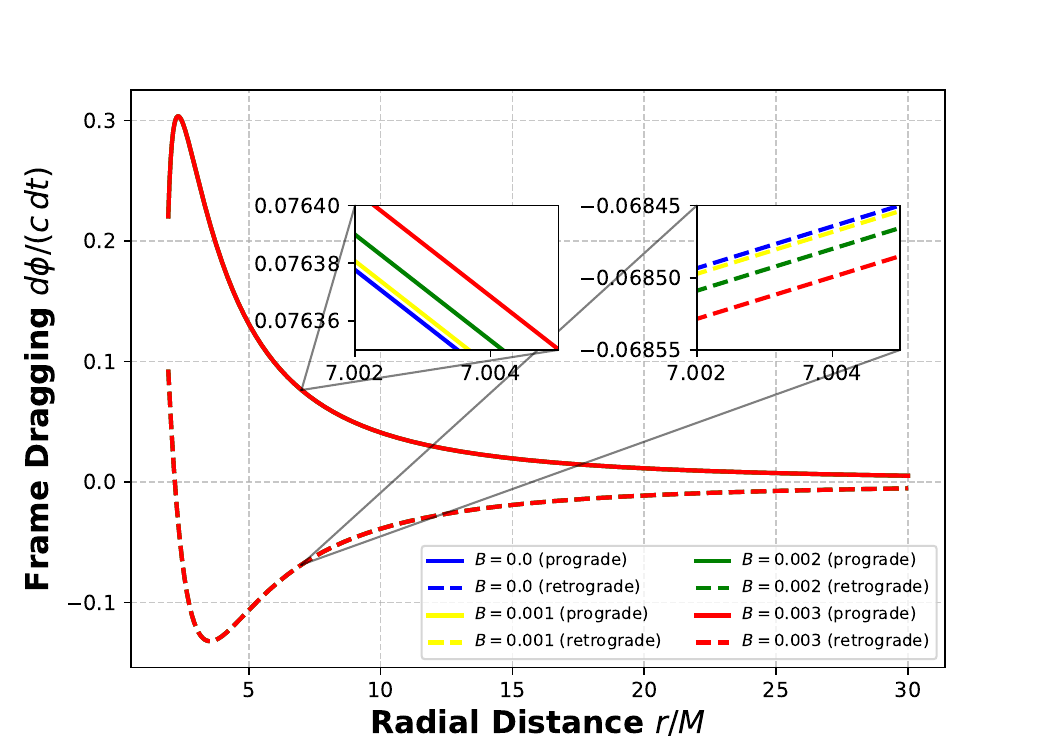}}
        \end{subfigure} 
        \begin{subfigure}[]
            {\includegraphics[width=80mm,height=60mm]{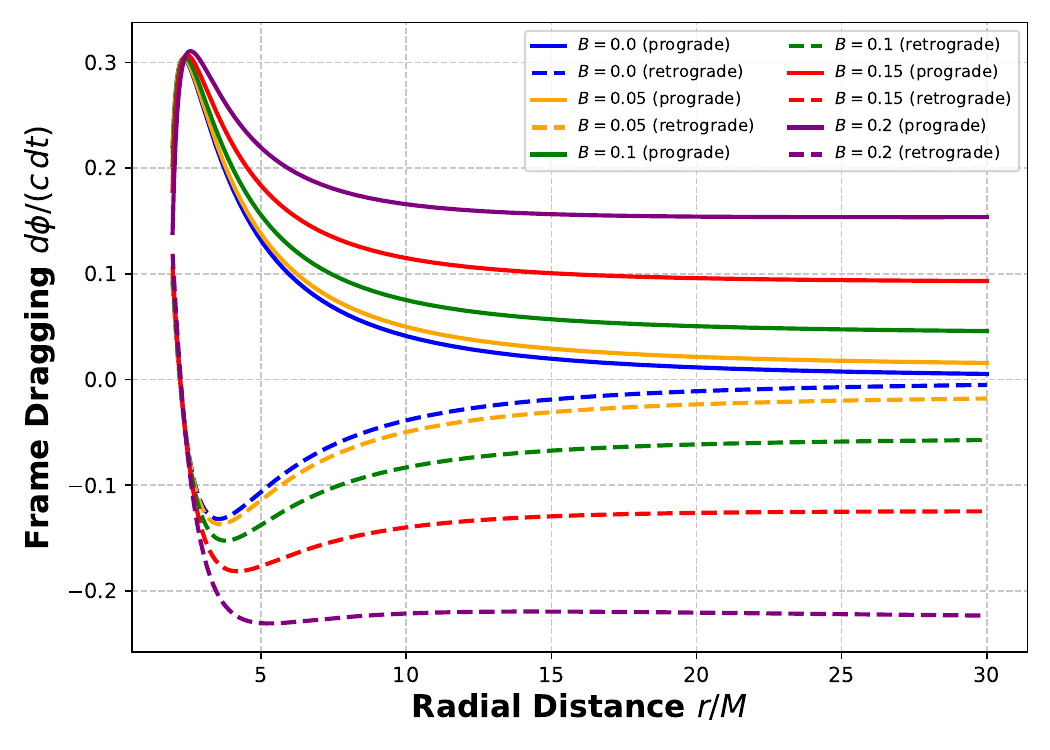}}
        \end{subfigure}
        \begin{subfigure}[]
            {\includegraphics[width=80mm,height=60mm]{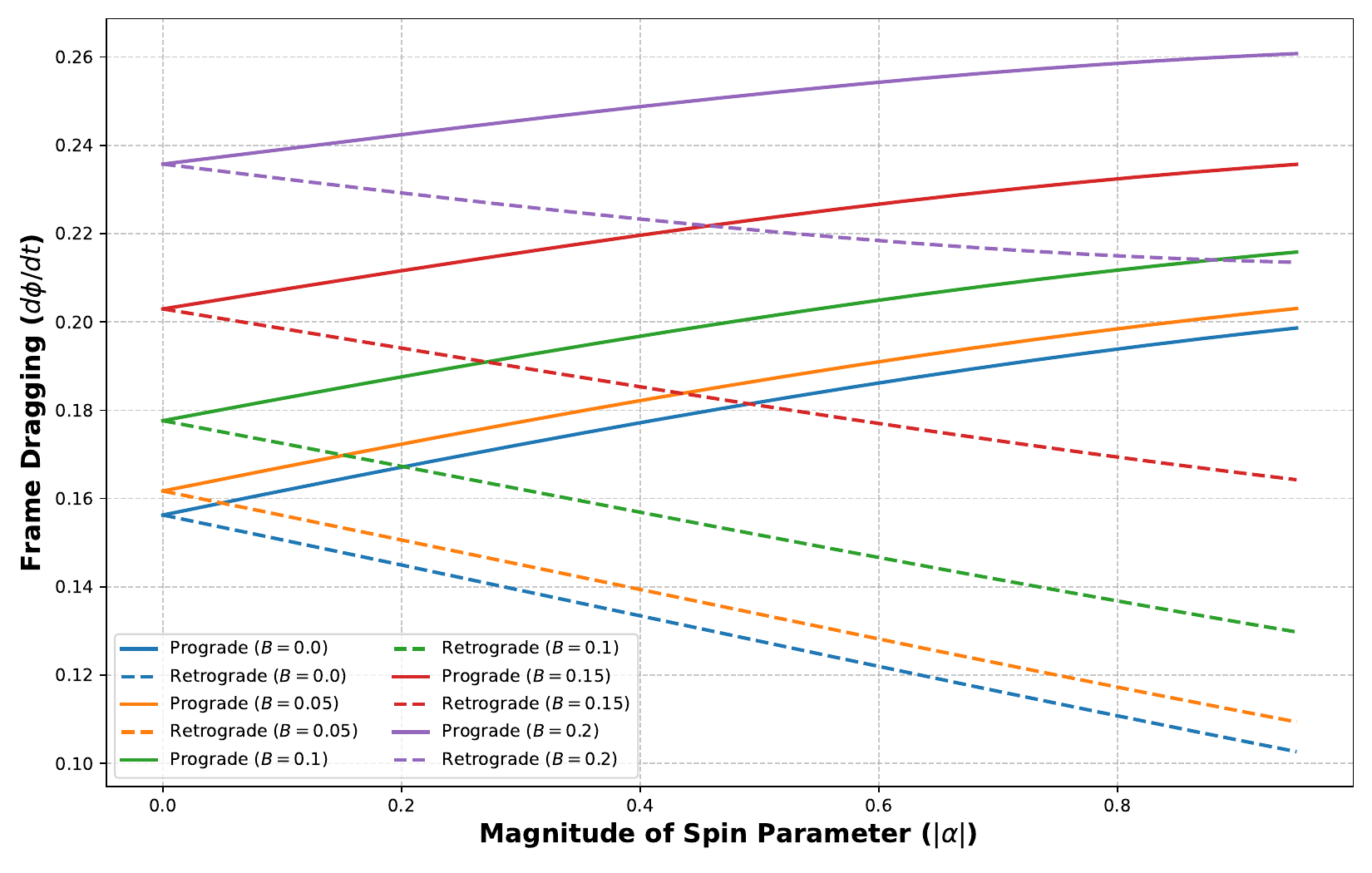}}
        \end{subfigure}
    \caption{Variation of frame dragging of KBR BH with (a) $\&$ (b) radial distance ($r$); (c) rotation parameter ($\alpha$); for different  values of magnetic field ($B$). Here, we consider $m=1$, $r=4M$, $\beta=5.0$ and $\alpha=0.5$ where we kept those parameters as constant.}
    \label{fig:frame dragging}
\end{figure}
Fig. \ref{fig:frame dragging} reflects the frame dragging effects due to KBR BH. Figs.~\ref{fig:frame dragging}~(a) and (b) map the frame dragging effect for traveling photons moving radially outward. Because these photons possess their own directional momentum, this represents a non-ZAMO case. This behaves completely differently from a Zero Angular Momentum Observer~(ZAMO). A ZAMO is forcefully dragged by the spacetime to a single positive peak near the BH before plunging into a negative, counter-rotating drag far away. In stark contrast, these traveling photons maintain their positive momentum and eventually stabilize into flat, horizontal plateaus rather than crossing below zero. Near the BH, the paths immediately split based on the photon’s direction. The retrograde curves (dashed) begin at a high angular velocity and drop steeply, while the prograde curves (solid) are heavily suppressed, climbing to a peak before falling. This massive initial gap proves that a prograde photon pushes through a significantly denser optical path. Facing higher optical resistance, its initial movement is severely restricted compared to a retrograde photon. The structural impact of the external magnetic field is visible at a long distance from the BH. For an unmagnetized BH (B = 0.0), the photon paths continuously decay toward zero, as the rotational drag fades in the empty space. However, as the magnetic field increases from 0.001 to 0.003 (in fig.~(a)) and 0.05 to 0.2 (in fig.~(b)), the curves shift higher and flatten into permanently elevated horizontal baselines. This visible flattening proves that the uniform magnetic background actively overpowers the localized spin at long distances, imposing a constant, permanent optical drag rather than allowing it to decay. In fig.(b) we have used exaggerated value of the magnetic field for clear visibility and differentiation of various curves at higher radial distances where as in fig.~(a) due to weak magnetic field all the curves are clearly visible only through inset.  
Fig.~\ref{fig:frame dragging}~(c) show how the BH spin (magnitude of the rotation parameter $|\alpha|$) changes the frame dragging effect. When the spin is zero ($|\alpha| = 0.0$), the prograde and retrograde lines for any specific magnetic field start at the exact same point. This visually proves that, without spin, the direction in which the photon moves makes no difference. As the spin increases, the lines clearly split in straight paths. The prograde lines (solid) slope upwards and the retrograde lines (dashed) slope downwards. This shows that the BH spin actively increases the dragging effect for a prograde photon and decreases it for a retrograde photon. Finally, at the different magnetic field strengths, we can see that as the magnetic field gets stronger (from B = 0.0 to B = 0.2), the entire pair of lines shifts vertically higher up on the graph. This visually confirms that the magnetic field directly raises the overall frame-dragging speed, acting as a constant boost that increases the dragging effect regardless of how fast the BH is spinning. 
\begin{figure*}[t]
\centering

\subfigure[]{
\includegraphics[width=80mm,height=71mm]{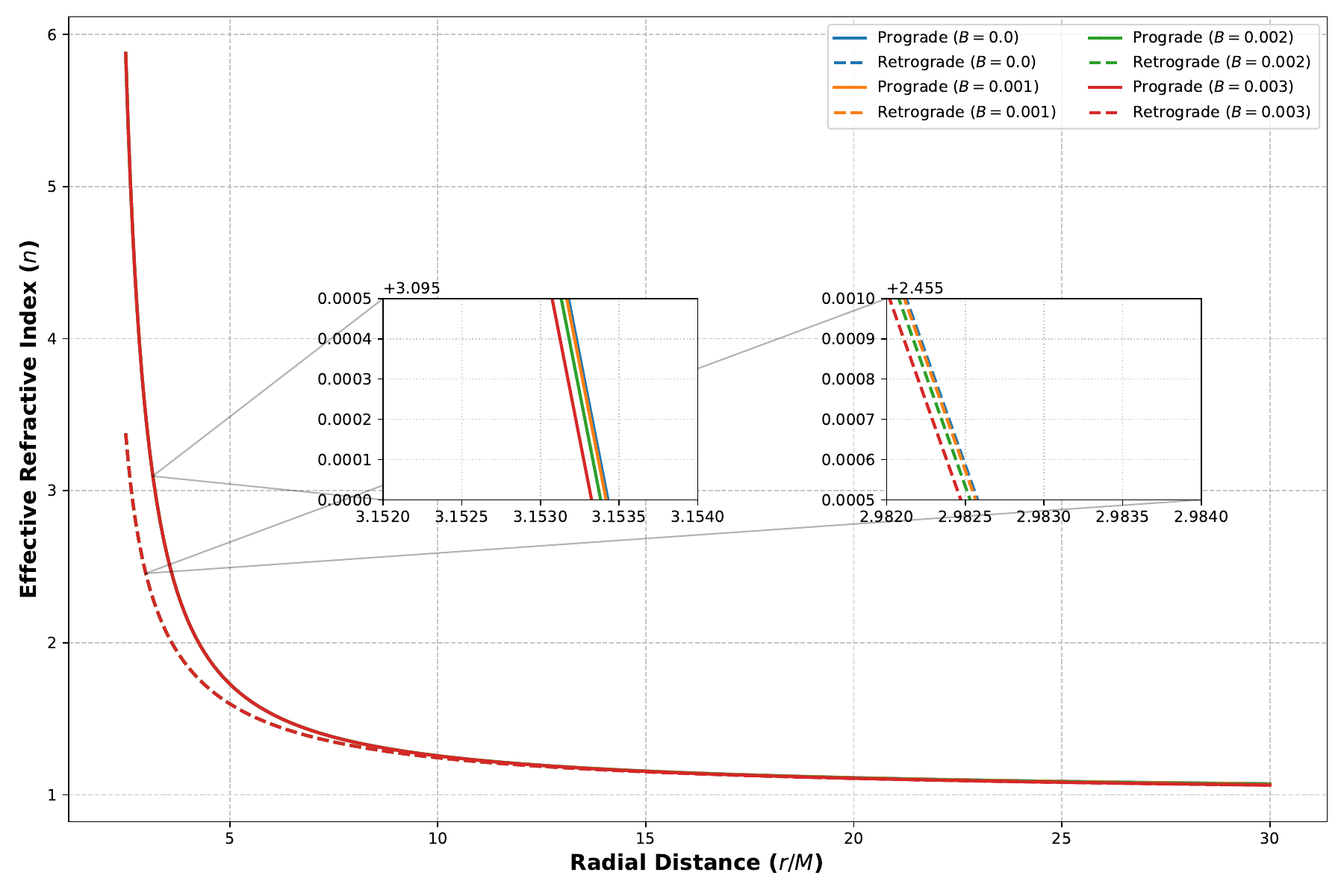}
}
\hfill
\subfigure[]{
\includegraphics[width=80mm,height=71mm]{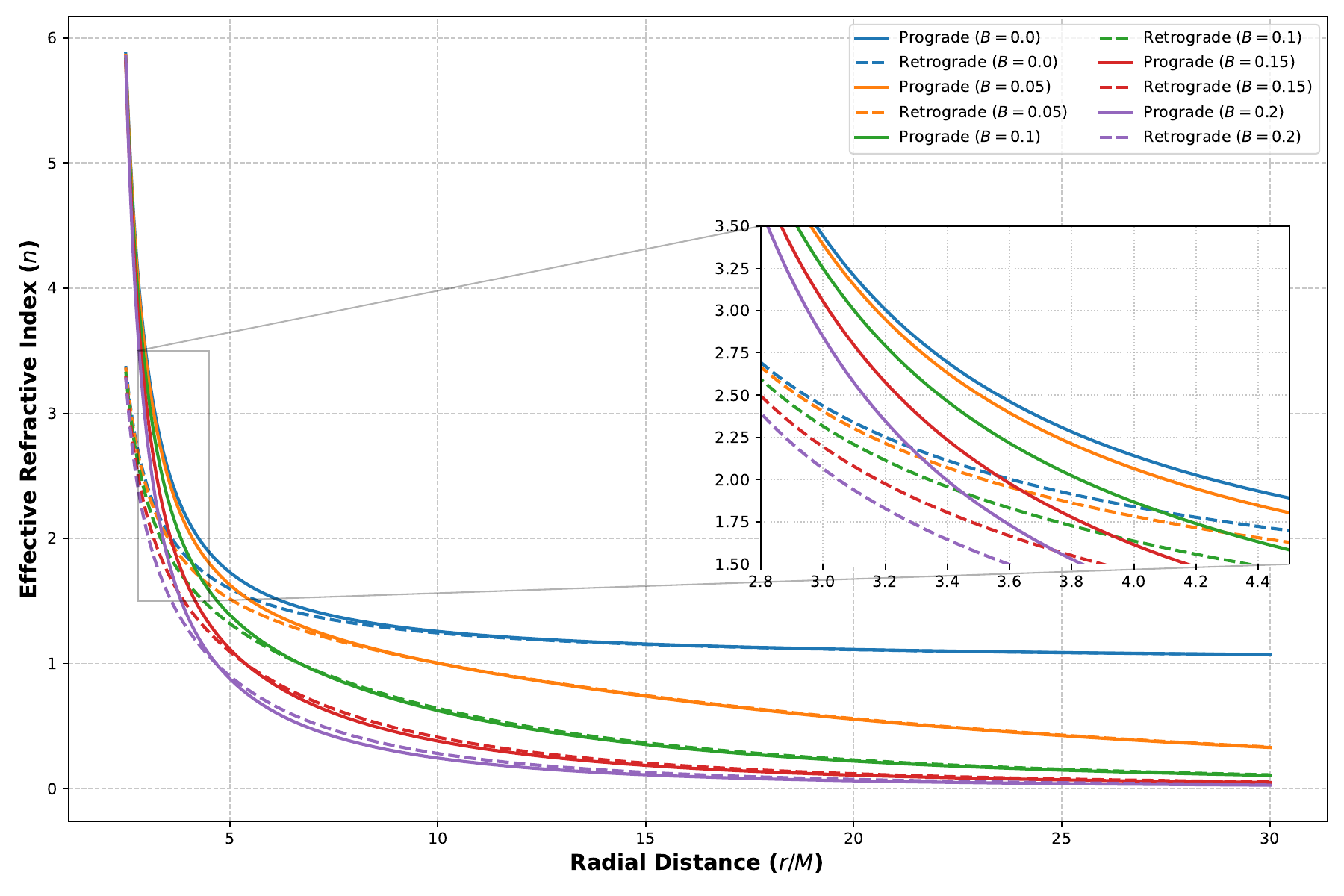}
}

\vspace{0.4cm}

\subfigure[]{
\includegraphics[width=80mm,height=71mm]{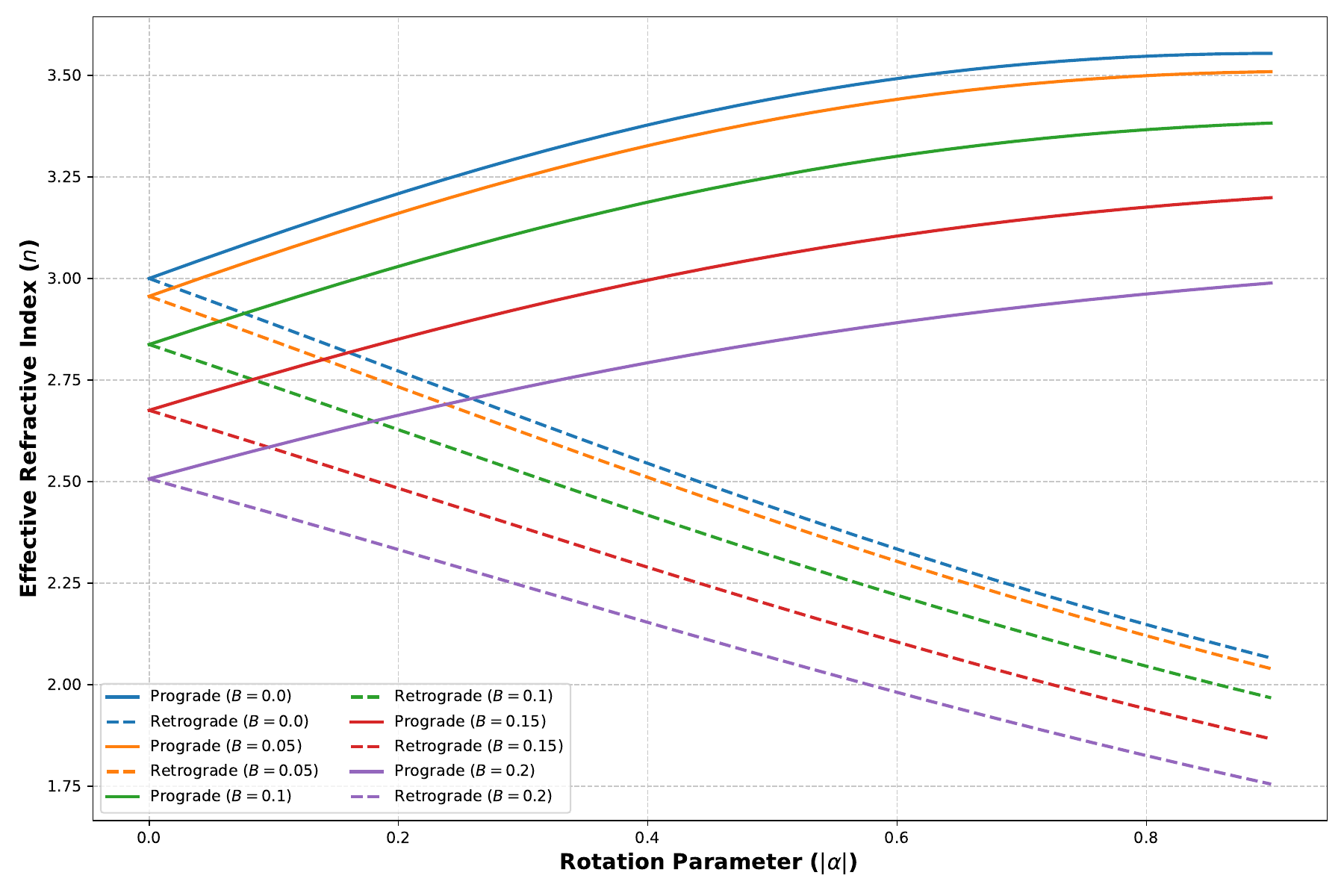}
}
\hfill
\subfigure[]{
\includegraphics[width=80mm,height=71mm]{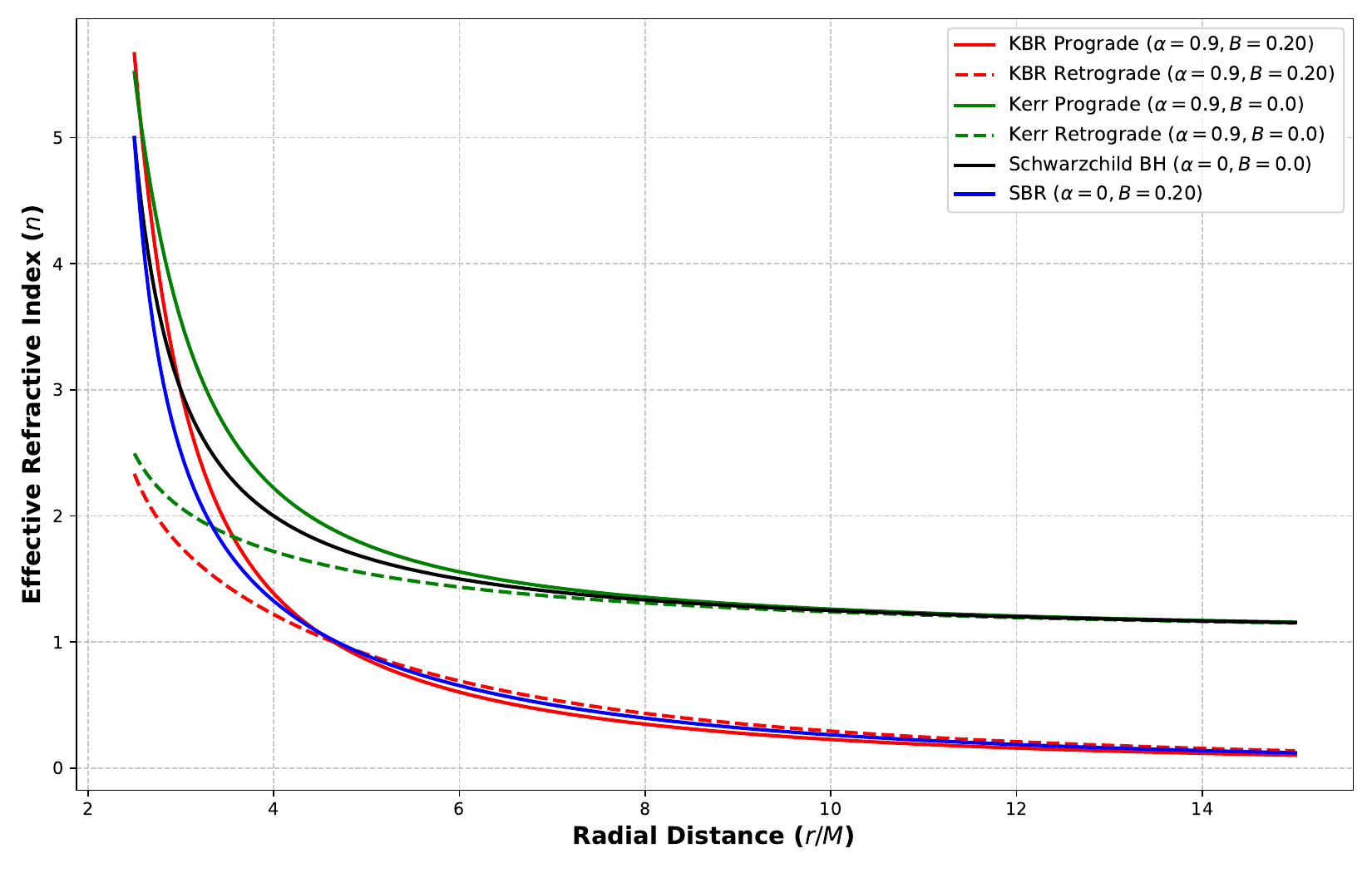}
}

\caption{
Variation of the refractive index of the KBR BH with (a) radial distance ($r$); (b) rotation parameter ($\alpha$) for different values of the magnetic field ($B$); (c) comparative study with different spacetimes; and (d) variation with the magnetic field $B$. Unless otherwise specified, the parameters are fixed at $m=1$, $r=3M$, and $\alpha=0.5$.
}
\label{fig:refractive index}
\end{figure*}
Fig.~\ref{fig:refractive index} reflects the behavior of the effective refractive index ($n$) as a function of radial distance ($r/M$)
and rotation parameter within the magnetized KBR spacetime, using the material medium approach. Figs.~\ref{fig:refractive index}~(a) and (b) clearly show how a fixed spin ($\alpha = 0.5$) disrupts the optical symmetry of the surrounding space. Near the BH, where gravitational effects are strongest, the refractive index is maximum indicating a very dense optical medium. As the photons move outward, the curves split into two separate paths. The prograde lines (solid) consistently stay higher than their corresponding retrograde lines (dashed). The detailed inset focusing on the near-horizon region shows that this gap confirms that a prograde photon must navigate a significantly denser optical path, facing more optical resistance than a retrograde photon.
Additionally, the external magnetic field ($B$) actively reshapes this optical medium. By comparing the baseline unmagnetized curves ($B = 0.0$) with the magnetized cases (up to $B = 0.20$), we observe that the magnetic field does not simply disappear in empty space. Instead, it systematically changes the refractive index at all distances, preventing the curves from settling into a Minkowski like flat vacuum. Moreover, as the magnetic field increases from 0.001 to 0.003 (in Fig.~\ref{fig:refractive index}~(a)) and 0.05 to 0.2 (in Fig.~\ref{fig:refractive index}~(b)), the curves shift lower.
Fig.~\ref{fig:refractive index}~(c) shows how the spin of a magnetized KBR BH (represented by the rotation parameter $|\alpha|$) affects the effective refractive index at a fixed radial distance close to the event horizon, specifically at $r = 3M$. When the spin is zero, that is, $|\alpha| = 0.0$, both the prograde and retrograde curves for any specific magnetic field strength start from the same point on the y-axis. This clearly indicates without rotation, the optical medium is completely symmetric. A photon encounters the same baseline optical resistance regardless of its path.
As the spin increases the curves noticeably diverge. This gap shows that rotation disrupts the symmetry of the optical medium. The faster the BH spins, the greater the difference between the two paths, leading to different scales of optical resistance based on whether the photon is moving against or with the flow of space time's rotational drag. The important role of the external magnetic field $B$ which defines the KBR spacetime, is evident when comparing the different sets of curves. As the magnetic field strength goes up from $B = 0.0$ to $B = 0.20$, the effective refractive index lowers for each pair of lines. This demonstrates that the magnetic field does not just remain in the background; it serves as a constant source of optical density, causes the overall refractive index of the surrounding space, creating a denser optical baseline even before considering the BH’s rotational effect.
Fig.~\ref{fig:refractive index}~(d) shows the basic optical behavior of the four different space times, helps us understand exactly what the rotation ($\alpha$) and the magnetic field ($B$) are doing. For the non-rotating cases (Schwarzschild and SBR), the optical medium is the same in all directions shows only a single line. But for the rotating cases (Kerr and KBR), the lines split into two: prograde (solid) and retrograde (dashed). This split proves that the BH rotation breaks the symmetry. Because of this, a photon fighting the spin faces a higher optical resistance. If we look closely at the magnetized KBR curves, we see a clear crossing point. Near the BH, the prograde line is above the retrograde line. But as we move further away, these two lines cross, and the prograde line falls below the retrograde line. The crossing point is the point where the frame-dragging of the BH gets overpowered by the strong magnetic background. After this point, the magnetic universe takes control and completely flips the optical rules. Finally, looking far away from the BH shows us the real effect of the magnetic field. The curves with no magnetic field (Kerr and Schwarzschild) flatten out perfectly at $n = 1.0$. This means they act like a normal, flat vacuum at large distances. But the magnetized curves (SBR and KBR) do not do this; they keep dropping towards $n = 0.0$. This proves that the magnetic field ($B$) permanently changes the space far away, preventing it from ever acting like a normal flat vacuum.

\section{Deflection due to Kerr-Bertotti-Robinson Black Hole} \label{sec4}
In the material medium approach, we treat the propagation of light in curved spacetime as motion through an equivalent optical medium. In such a case, the gravitational field is represented by an effective refractive index $n(r)$, allowing the trajectory of light rays to be determined using the principles of geometrical optics. In this framework, the curvature of spacetime is directly encoded in the spatial variation of this refractive index.

The following expression represents the deflection integral in terms of refractive index used to calculate the total deflection angle~\cite{born2013principles}:

\begin{equation}
\Delta \phi = 2 \int_{\beta}^{\infty}
\frac{dr}{r \sqrt{\left( \frac{n(r)\, r}{n(\beta)\, \beta} \right)^2 - 1}} - \pi ,
\end{equation}
where $\beta$ represents the actual minimum
physical distance or impact parameter. In the present case, light approaches from asymptotic infinity, parallel to the x-axis ($r =- \infty$) toward a charged rotating gravitational mass at the origin, defined by the Schwarzschild radius $r_g$, rotation parameter $\alpha$, and external magnetic field $B$.
The beam reaches $r = \infty$ after a specified amount of deflection ($\Delta\psi$). The closest approach distance to the incoming ray (known as the impact parameter) is the perpendicular distance between the center of the KBR BH and the ray. 

To obtain a solvable analytical expression for the deflection angle and check how rotation parameter $\alpha$, and external magnetic field $B$ are affected, we must apply some extensive algebraic manipulation on the exact refractive index expression derived earlier in Eqs.~(\ref{Eq_ref}) or (\ref{Eq_ref_final}). (shown in appendix A) 
So, with this extensive work the refractive index becomes
\begin{eqnarray}\label{refrac_manipul}
n(r) &=&\frac{1}{a} \left[ 1 + \frac{b}{a r} - B^2 r^2 - \alpha Y(r)\frac{d\phi}{cdt} \right]\nonumber\\
&=& \frac{1}{a} \left[ 1 + \delta (r) \right],
\end{eqnarray}
where 
{\small
\begin{eqnarray*}
    a&=&\left(1-\frac{B^2 I_2 m^2 }{I_1^2} \right),\\
    b&=&\frac{2 I_{2} m }{ I_{1} },\\
    Y(r)& =& (\frac{1}{a} - 1) + \frac{b}{a^2 r} - \frac{B^2 r^2}{a},\\
    \delta (r) &=& \frac{b}{a r} - B^2 r^2 - \alpha Y(r)\frac{d\phi}{cdt}.
\end{eqnarray*}
Using Eq.~(\ref{refrac_manipul}), we obtain
the expression for the trajectory as
\begin{eqnarray}\label{def_final}
    \Delta \phi &=& 2 \int_{\beta}^{\infty}
\frac{dr}{r \sqrt{\frac{r^2}{\beta^2}\big(1+\delta(r)-\delta(\beta)\big)^2-1}} - \pi\nonumber\\
&=& 2 \int_{\beta}^{\infty}
\frac{dr}{r \sqrt{\frac{r^2}{\beta^2}\big(1+2[\delta(r)-\delta(\beta)]\big)-1}} - \pi\nonumber\\
&=& 2 \int_{\beta}^{\infty}\frac{dr}{r \sqrt{ \left(\frac{r^2}{\beta^2} - 1\right) + \frac{r^2}{\beta^2} \cdot 2(\delta(r) - \delta(\beta)) }}-\pi\nonumber\\
&=& 2 \int_{\beta}^{\infty}\frac{1}{r \sqrt{\frac{r^2}{\beta^2} - 1}} \left[ 1 + \frac{r^2}{\beta^2} \cdot \frac{2(\delta(r) - \delta(\beta))} {\frac{r^2}{\beta^2} - 1} \right]^{-1/2}dr-\pi\nonumber\\
&=& 2 \int_{\beta}^{\infty}\frac{\beta}{r \sqrt{r^2-\beta^2}} \left[ 1 - r^2 \cdot \frac{\delta(r) - \delta(\beta)} {r^2 - \beta^2} \right] dr -\pi.
\end{eqnarray}
}
The final expression~(\ref{def_final}) gives the total angular deflection in the KBR spacetime. This bending is not just caused by the central mass $m$; it is strongly influenced by the rotation $\alpha$ and the external magnetic field $B$. Because the magnetic field alters the far-field refractive index, a KBR BH lens deflects the light differently than a standard Kerr BH. Evaluating this integral for specific impact parameters $\beta$ allows us to map exactly how the magnetic environment modifies this deflection, completing our material medium analysis.

\begin{figure*}[t]
\centering

\subfigure[]{
\includegraphics[width=85mm,height=75mm]{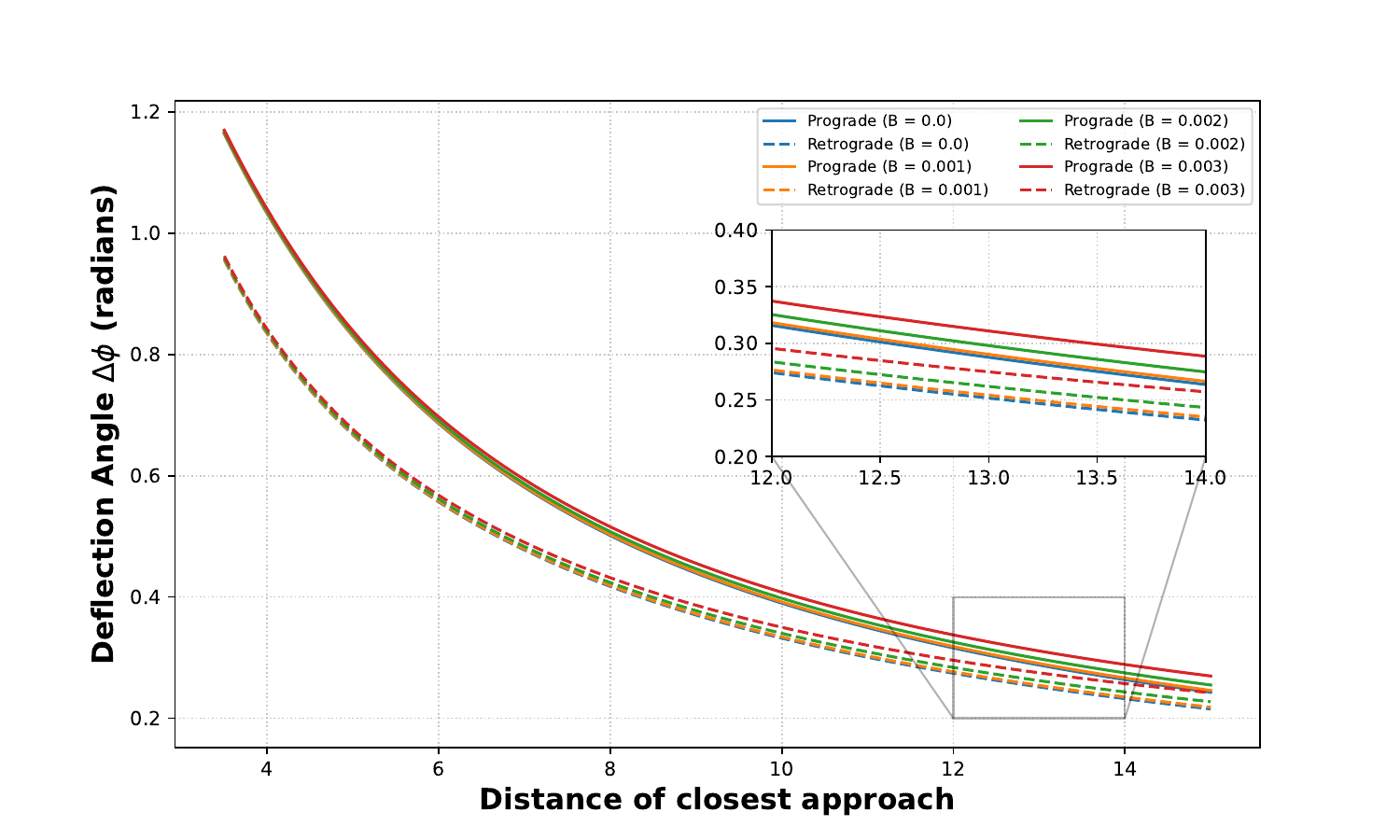}
}
\hfill
\subfigure[]{
\includegraphics[width=80mm,height=70mm]{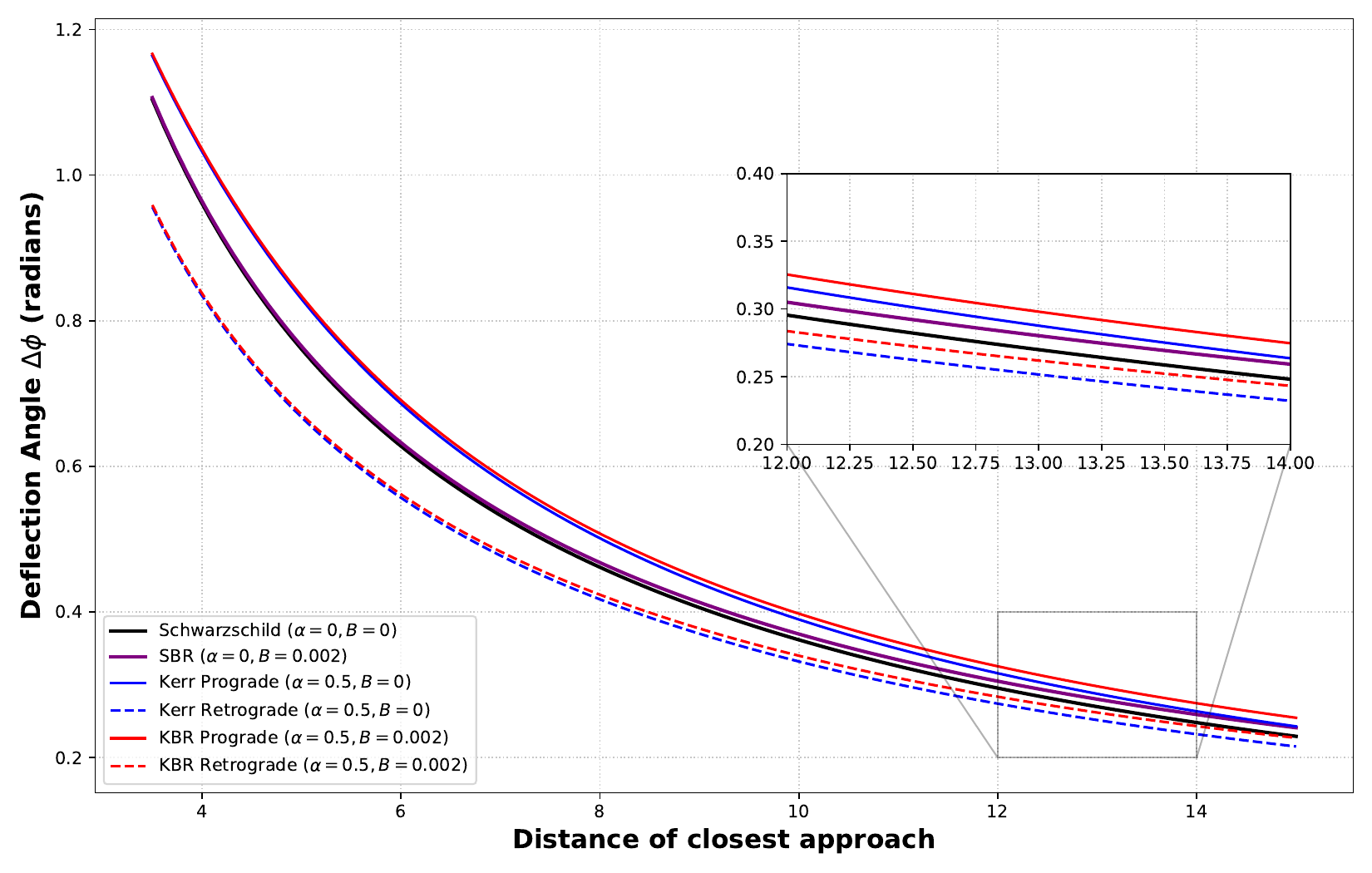}
}

\vspace{0.4cm}

\subfigure[]{
\includegraphics[width=80mm,height=70mm]{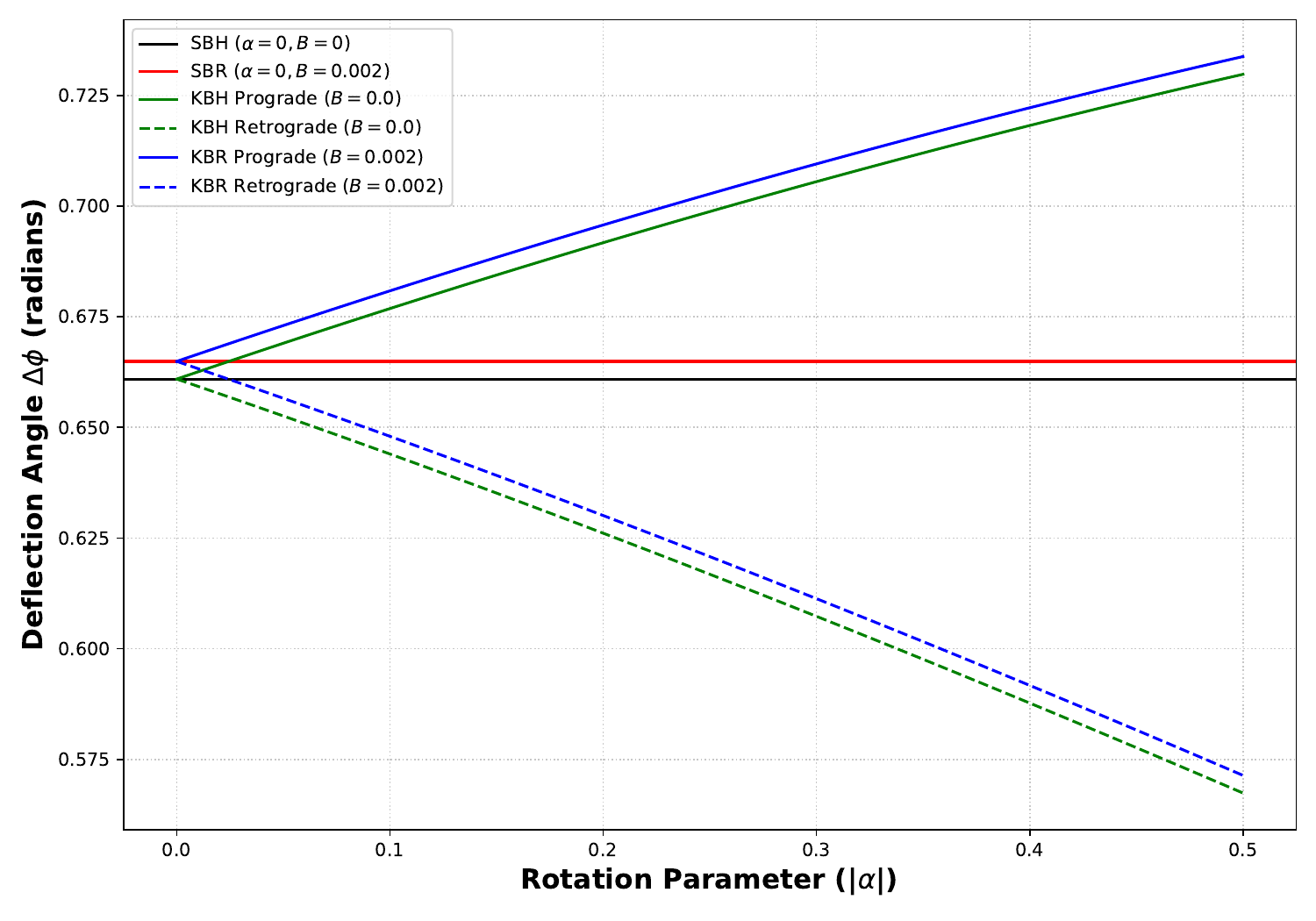}
}
\hfill
\subfigure[]{
\includegraphics[width=80mm,height=70mm]{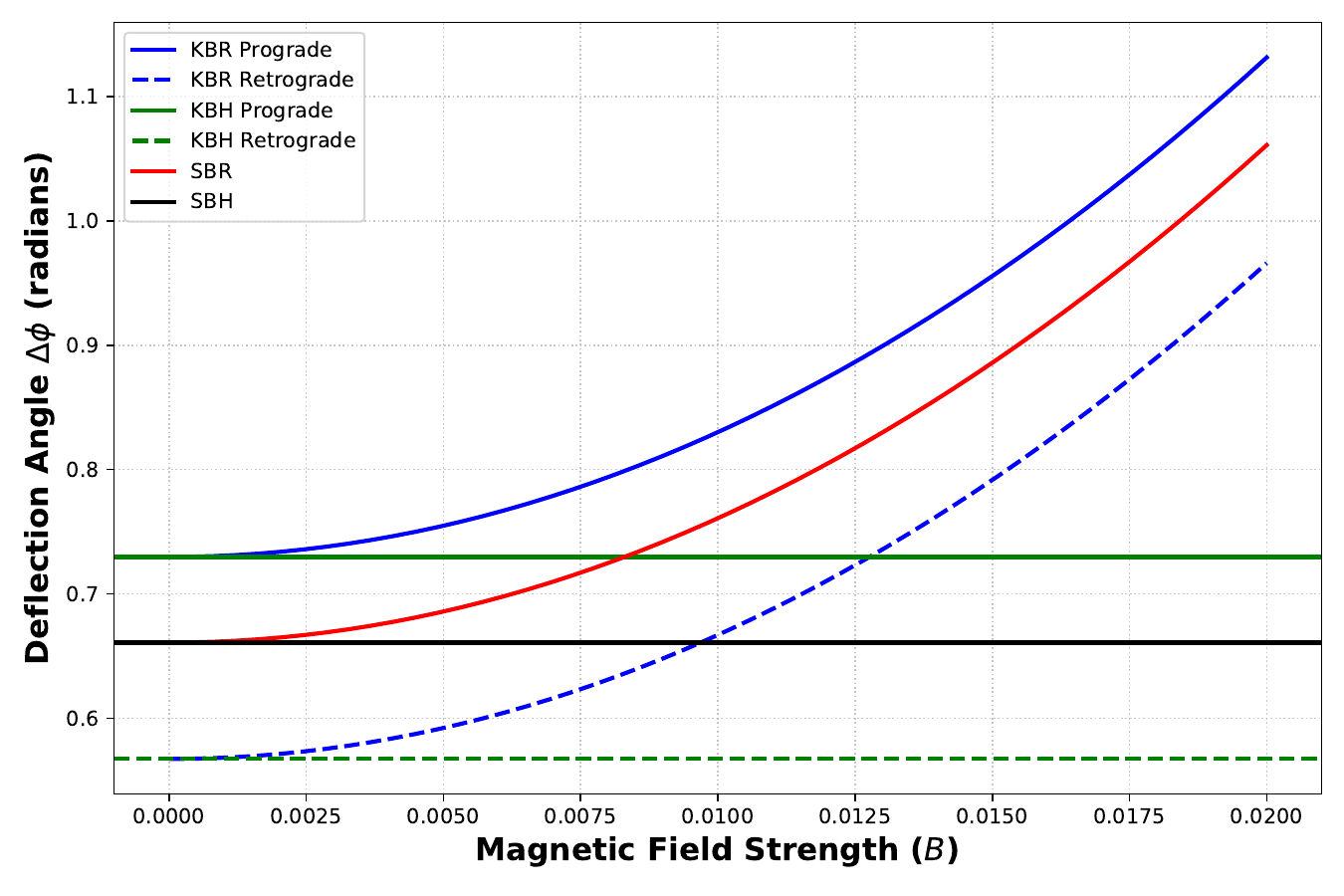}
}

\caption{Variation of deflection angle of KBR BH with (a) radial distance ($r$) at different magnetic field ($B$). A Comparative study of deflection angle with different spacetime with (b) radial distance ($r$); (c) magnetic field ($B$); (d) rotation parameter ($\alpha$). Here, we consider $m=1$, $r=5M $ and $\alpha=0.5$ where we kept those parameters as constant.}
\label{fig:deflection angle}
\end{figure*}
Fig.~\ref{fig:deflection angle}~(a) shows the variation of the deflection angle of the KBR BH with the radial distance ($r$) at different magnetic fields ($B$). In addition, a comparative study of the deflection angle with different spacetimes is carried out with radial distance ($r$) in Fig.~\ref{fig:deflection angle}~(b); with rotation parameter ($\alpha$) in Fig.~\ref{fig:deflection angle}~(c); and with magnetic field ($B$) in Fig.~\ref{fig:deflection angle}~(d).

The plot in Fig.~\ref{fig:deflection angle}~(a) shows how much light bends (the deflection angle) as the distance of closest approach or impact parameter $\beta$ increases in the KBR spacetime. It is seen that all the curves steadily drop downward, which means that the bending effect gets weaker the further away the photon passes by the BH, which is exactly what we expect as gravity drops off. Another consistent gap between the prograde and retrograde lines is observed for every single magnetic field strength. The prograde lines (solid) always stay above their matching retrograde lines (dashed) that proves that a prograde photon has to push through a ``denser” optical path. Because it faces higher optical resistance, it ends up bending significantly more than a retrograde photon. Finally, the direct effect of the magnetic field may be concluded by looking at the different $B$ values in the legend. As the magnetic field increases from 0.0 to 0.003 (this small interval is due to the weak magnetic field assumption), the entire pair of lines shifts higher up on the graph, showing that the magnetic field actively adds to the BH’s gravity, making the overall bending of light even stronger.
Fig.~\ref{fig:deflection angle}~(b) compares how strongly each of the four different spacetime bends light. Since the Schwarzschild and SBR spacetimes do not rotate ($\alpha = 0$), they only have a single line, which acts as a neutral baseline. The rotating Kerr and KBR spacetimes ($\alpha = 0.5$) split perfectly above and below these baselines. This clearly shows that rotation creates a split effect: it increases the bending for prograde photons and decreases for retrograde photons. The most important physical detail to notice is the placement of the KBR prograde line, which sits at the very top of the graph. Since this specific line is the only one that combines both rotation ($\alpha = 0.5$) and a magnetic field ($B = 0.002$), it visually proves that having both working together creates the strongest possible light bending among the cases shown.
Fig.~\ref{fig:deflection angle}~(c) shows how the spin of the BH directly changes the bending of light. In non-rotating cases (SBH and SBR), these are perfectly flat, horizontal lines, visually makes perfect sense because they do not have any spin so that along the rotation axis has zero effect on them. The rotating cases both start at the exact same points as those flat baselines when the spin is zero: the unmagnetized KBH lines start exactly on the SBH baseline, while the magnetized KBR lines start exactly on the higher SBR baseline. This SBR baseline sits higher because the magnetic field ($B = 0.002$) increases the amount of bending even before any spin is added. As the spin increases moving to the right, the rotating lines clearly split apart. The prograde lines slope upwards, and the retrograde lines slope downwards. By looking at this widening gap, it visually proves a simple rule: the faster the BH spins, the larger the difference becomes between the two photon paths. The spin actively forces them away from the baseline, increasing the bending for the prograde photon and decreasing for the retrograde one. Finally, by comparing the two rotating setups, we can clearly see the combined effect. The entire set of magnetized KBR lines ($B = 0.002$) sits completely above the unmagnetized KBH lines ($B = 0.0$). This visually confirms a clear division of roles: while the rotation parameter is what creates the growing split, the magnetic field is what raises the overall baseline amount of bending.
Fig.~\ref{fig:deflection angle}~(d) looks specifically at how changing the magnetic field strength ($B$) affects the bending of light. The lines representing the spacetime with zero magnetic field (SBH and both KBH lines) are perfectly flat and horizontal confirms that because they do not have a magnetic field and hence no effect on them. The non-rotating SBR line curves smoothly upward, which shows the baseline effect of a magnetic field adding to the bending of light. The rotating KBR lines split clearly around this SBR baseline. The KBR prograde line stays well above the SBR line, while the KBR retrograde line stays below it, proves that rotation creates a strict split in how the magnetic field affects the photons across all tested magnetic strengths. A prograde photon will always experience more bending than the non-rotating baseline, and a retrograde photon will always experience less.
\section{Thermodynamics of Kerr-Bertotti-Robinson BH} \label{sec5}
Although BHs are considered massive purely gravitational objects, they also have deep  connections to the field of thermodynamics~\cite{Bekenstein73,Hawking75}. Bekenstein and Stephen Hawking showed that the surface area of BH event horizon and surface gravity are strikingly linked  with thermodynamic properties such as entropy and temperature. 
\subsection{Surface Area of Event Horizon}
The surface area of the event horizon of the BH is directly linked to the Entropy of Bhs through the Bekenstein Hawking relation~\cite{Bekenstein73}. For an axis-symmetric  stationary BH, like KBR BH, the horizon area is computed by integrating over the two-dimensional spatial cross-section of the event horizon as
\begin{equation}
A = \int_{0}^{2\pi} \int_{0}^{\pi} \sqrt{g_{\theta\theta}\, g_{\phi\phi}} \, d\theta \, d\phi.
\end{equation}
For the KBR metric the surface Area is given by 
\begin{equation}
A= \frac{4\pi({r_{H}}^2+\alpha^2)}{1+B^2{r_{H}}^2},
\end{equation}
where, ${r_H}$ is the radius at the event horizon, obtained in Eq.~\ref{horizon}. The presence of the magnetic field B changes the overall size of the horizon area compared to that of a normal Kerr BH.

\subsection{Entropy}
The entropy of a BH is proportional to the area of the event horizon rather than volume, highlighting a fundamental departure from conventional thermodynamic systems which was proposed by Jacob Bekenstein~\cite{Bekenstein73} and is given by the Bekenstein–Hawking relation
 \begin{equation}
     S=\frac{A}{4},
 \end{equation}
where A is the area of the event horizon, expressed in natural units. Using the Bekenstein–Hawking relation on the KBR BH, the entropy is found to be 
\begin{equation}
    S= \frac{\pi({r_H}^2+\alpha^2)}{1+B^2{r_H}^2}.
\end{equation}
Here again, we can see that the magnetic field $B$ directly affects the macroscopic state of the BH.

\begin{figure}
    \centering
        \begin{subfigure}[]
            {\includegraphics[width=80mm,height=60mm]{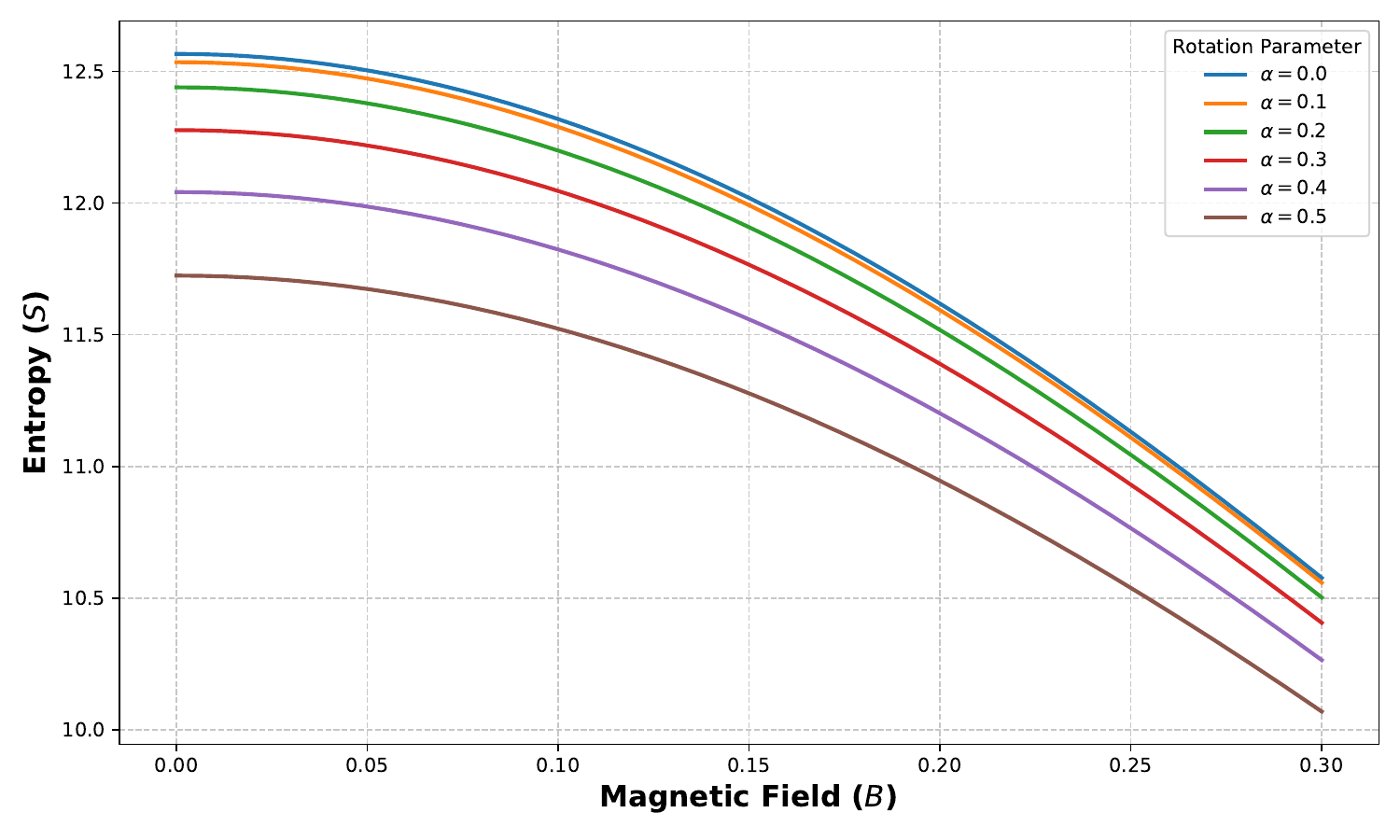}}
        \end{subfigure}
        \begin{subfigure}[]
            {\includegraphics[width=80mm,height=60mm]{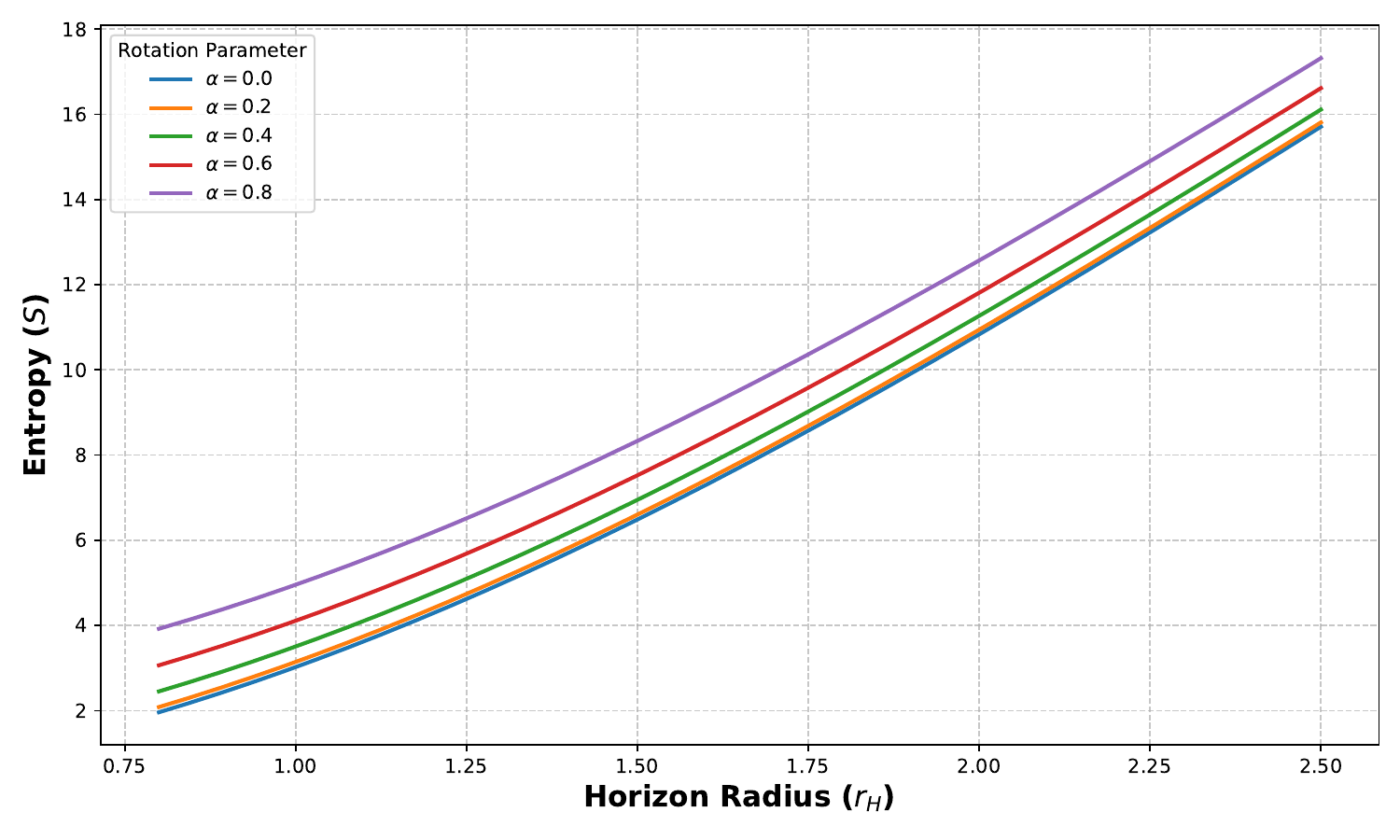}}
        \end{subfigure}
        \begin{subfigure}[]
            {\includegraphics[width=80mm,height=60mm]{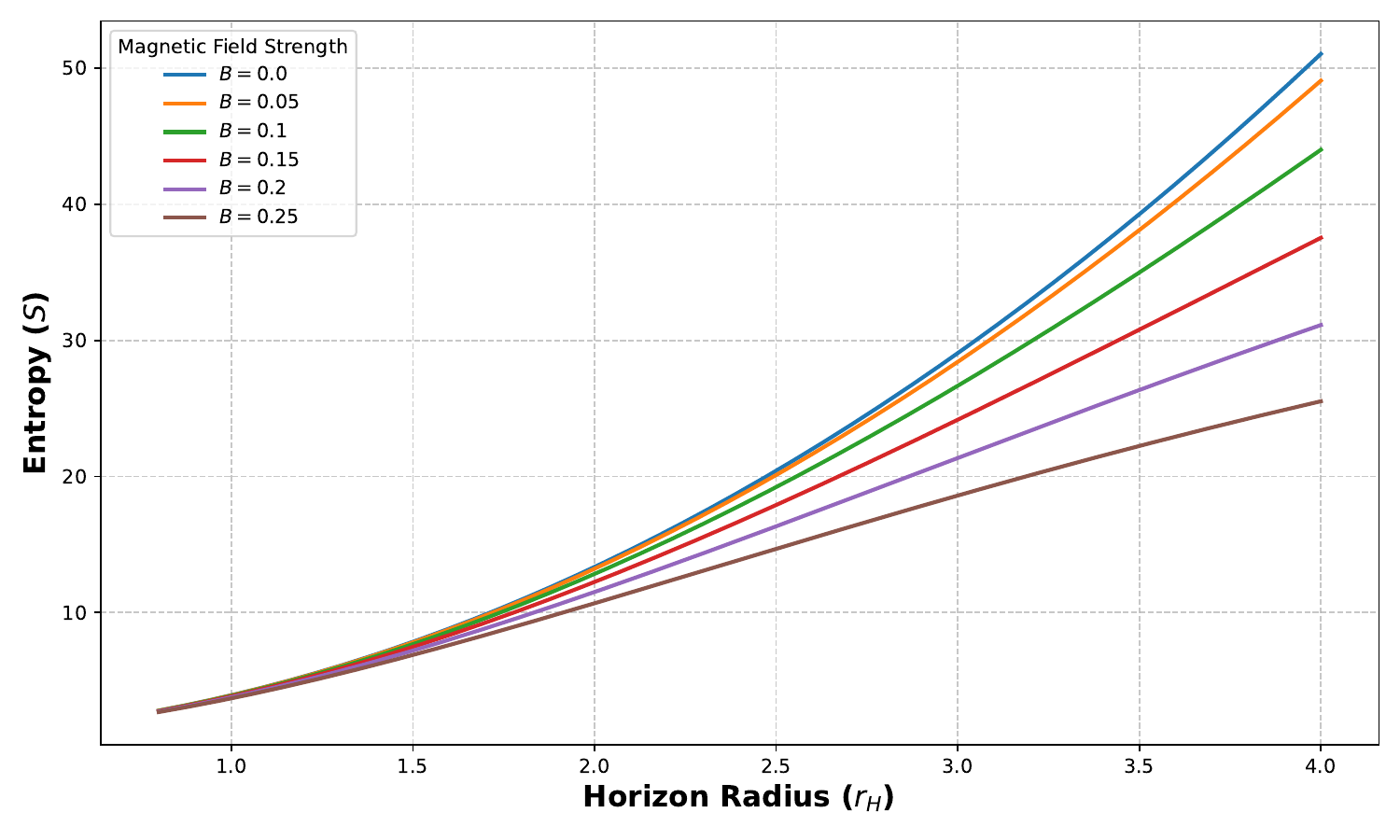}}
        \end{subfigure}
    \caption{Variation of Entropy ($S$) of KBR BH with (a) magnetic field ($B$) for different  values of rotation parameter ($\alpha$); (b) horizon radius ($r_H$) for different  values of rotation parameter ($\alpha$); (c) horizon radius ($r_H$) for different  values of magnetic field ($B$). Here, we consider $m=1$, $B=0.2$ and $\alpha=0.5$ where we kept those parameters as constant.}
    \label{fig:entropy}
\end{figure}
Fig.~\ref{fig:entropy} shows the graphical representation of the Bekenstein-Hawking entropy $S$ of KBR BH with magnetic field and horizon radius.
In Fig.~\ref{fig:entropy}~(a), we plot multiple curves corresponding to different rotation parameters ($\alpha$). The primary trend illustrates the profound impact of the external magnetic field $B$ on the macroscopic state of the BH: as the magnetic field strength increases, the entropy strictly decreases across all spin values. This demonstrates that the omnipresent electromagnetic background of the Bertotti-Robinson universe exerts a compressing effect on the geometry of the event horizon. 
Furthermore, the variation of $\alpha$ values highlights the role of intrinsic rotation. For any given magnetic field strength, a BH with a higher rotation parameter (faster spin) exhibits a lower overall entropy. This proves that both the external magnetic environment and the internal angular momentum work together to geometrically restrict the event horizon, thereby limiting the overall information-storing capacity of the BH compared to a static, unmagnetized vacuum equivalent.

To further isolate the thermodynamic behavior of the Kerr-Bertotti-Robinson metric, we treat the event horizon radius $r_H$ as an independent continuous variable in Figs.~\ref{fig:entropy}~(b) and (c). The plot of the Bekenstein-Hawking entropy $S$ against the horizon radius reveals a profound deviation from standard BH thermodynamics. In an unmagnetized vacuum, BH entropy scales purely quadratically with its radius as $S \propto r_H^2$. However, for the KBR spacetime, the presence of a uniform magnetic background $B$ introduces a quadratic term in the denominator. As the horizon expands, this magnetic term dampens the unbounded quadratic growth, forcing the entropy to scale much slower than a standard Kerr BH, implies that the omnipresent electromagnetic energy density of the Bertotti-Robinson universe fundamentally restricts the maximum information-storing capacity of spacetime as the BH grows. Furthermore, the curves for the varying rotation parameter $\alpha$ demonstrate that for any horizon size, intrinsic spin contributes to the geometric area, slightly elevating the baseline entropy.
Fig.~\ref{fig:entropy}~(c) illustrates how the Bekenstein-Hawking entropy $S$ changes as the horizon radius $r_H$ increases, keeping the rotation parameter fixed at $\alpha = 0.5$ while varying the magnetic field strength $B$. The fundamental trend is that, with the horizon radius, the entropy naturally increases, which aligns with the fact that a larger BH has a larger surface area. For an unmagnetized BH, $B=0$ and the entropy grows very rapidly. But as we introduce and increase the external magnetic field $B$, the curves become significantly flattened, which means that the uniform electromagnetic energy of the background universe actually compresses the geometry of the BH. This compression restricts how much surface area and information, or entropy, the BH can support for any given horizon size.
 
\subsection{Surface gravity}
Surface gravity is the acceleration required to hold a unit mass stationary at the event horizon of a BH as measured by an observer at infinity. This quantity plays an analogous role to temperature in BH thermodynamics and is a fundamental quantity in BH Mechanics as well~\cite{Bardeen73}.
For a stationary BH spacetime, surface gravity is defined in terms of the null Killing vector field that becomes null on the event horizon and is defined as
\begin{equation}
\xi^b \nabla_b \xi^a = \kappa \, \xi^a,
\end{equation}
where $\xi^a$ is the timelike Killing vector and $\kappa$ is surface gravity which is evaluated exactly at the horizon. 
For the KBR metric, $\kappa$ is given by 
\begin{equation}
    \kappa = \frac{\dot{Q}(r_H)}{2({r_H}^2 +\alpha^2)},
\end{equation}
where $\dot{Q}(r_H)$ is the derivative of the function $Q(r_H)$ with respect to $r$ and in KBR spacetime as
\begin{equation}
    \dot{Q}(r_H)=\frac{2(1+B^2{r_H}^2)({I_1}^2 r_H -m^2B^2r_HI_2  -2mI_1 I_2)}{{I_1}^2}.
\end{equation}
Thus, the expression for $\kappa$ becomes
\begin{equation}
    \kappa=\frac{(1+B^2{r_H}^2)({I_1}^2 r_H -m^2B^2r_HI_2  -2mI_1 I_2)}{{I_1}^2({r_H}^2 + \alpha^2)}.
\end{equation} 
\subsection{Hawking Temperature}
BHs emit thermal radiation due to quantum effects near the event horizon and are therefore said to have a temperature, called Hawking temperature~\cite{Hawking75}. The Hawking temperature is directly proportional to the surface gravity of the BH as follows:
\begin{equation}
    T=\frac{\kappa}{2\pi}.
\end{equation}
This relation establishes a connection between gravity, thermodynamics, and quantum mechanics. Through this equation, we can show that BHs can lose mass over time and possibly evaporate. It also shows that the smaller the BH is, the hotter it is, and the larger it is, the colder it becomes, indicating that large BHs have a longer lifespan.
  
By substituting our analytical expression for the surface gravity $\kappa$, we are able to arrive at the final expression for the Hawking Temperature for the KBR Metric: 
\begin{equation}
    T=\frac{(1+B^2{r_H}^2)({I_1}^2 r_H -m^2B^2r_HI_2  -2mI_1 I_2)}{2\pi{I_1}^2({r_H}^2 +\alpha^2)}.
\end{equation}
This equation tells us that the temperature of this BH does not just depend on its mass and rotation ($\alpha$), but is heavily influenced by the uniform electromagnetic background ($B$). The interaction between the rotation and the magnetic field completely changes how the BH behaves thermodynamically.

\begin{figure}
    \centering
        \begin{subfigure}[]
            {\includegraphics[width=80mm,height=60mm]{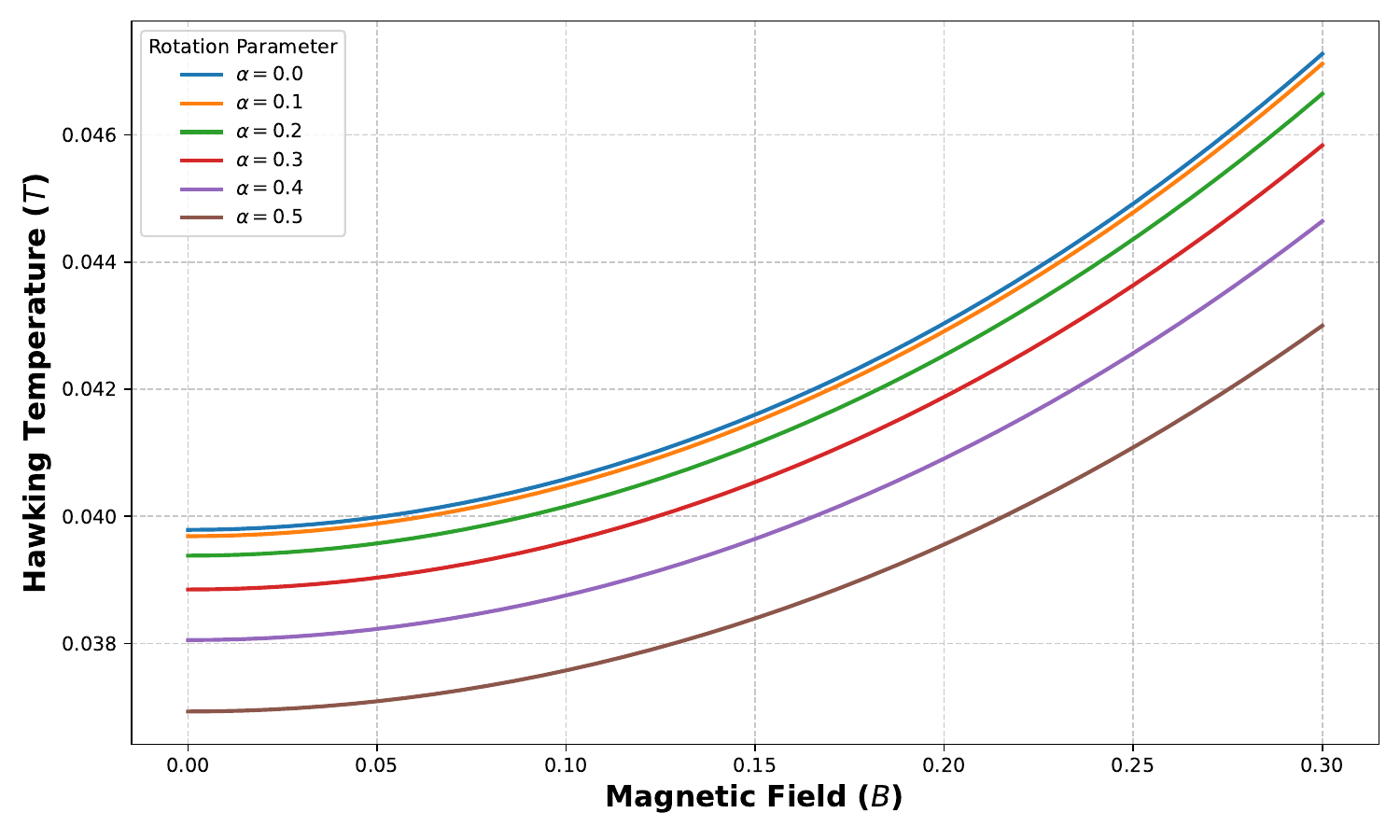}}
        \end{subfigure}
        \begin{subfigure}[]
            {\includegraphics[width=80mm,height=60mm]{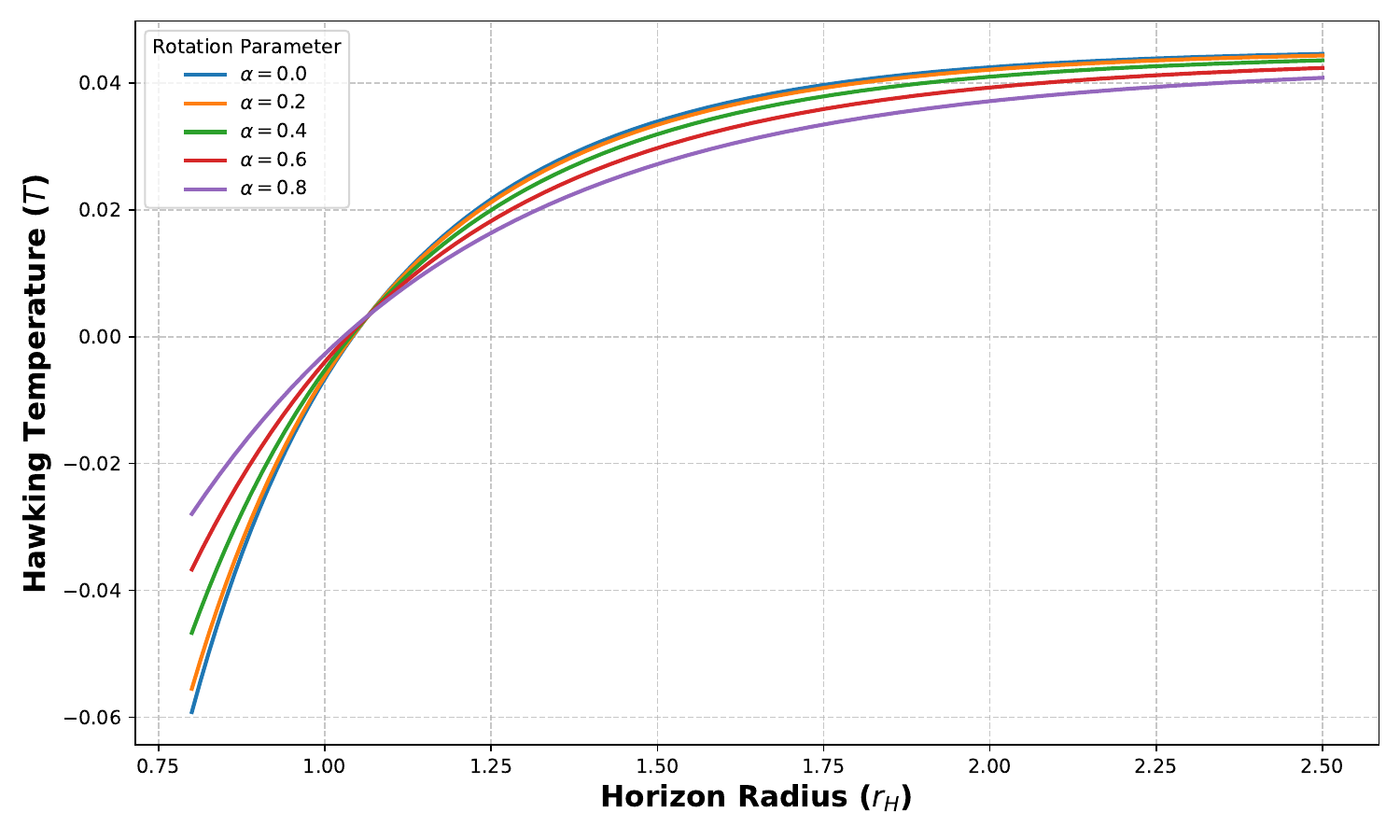}}
        \end{subfigure}
        \begin{subfigure}[]
            {\includegraphics[width=80mm,height=60mm]{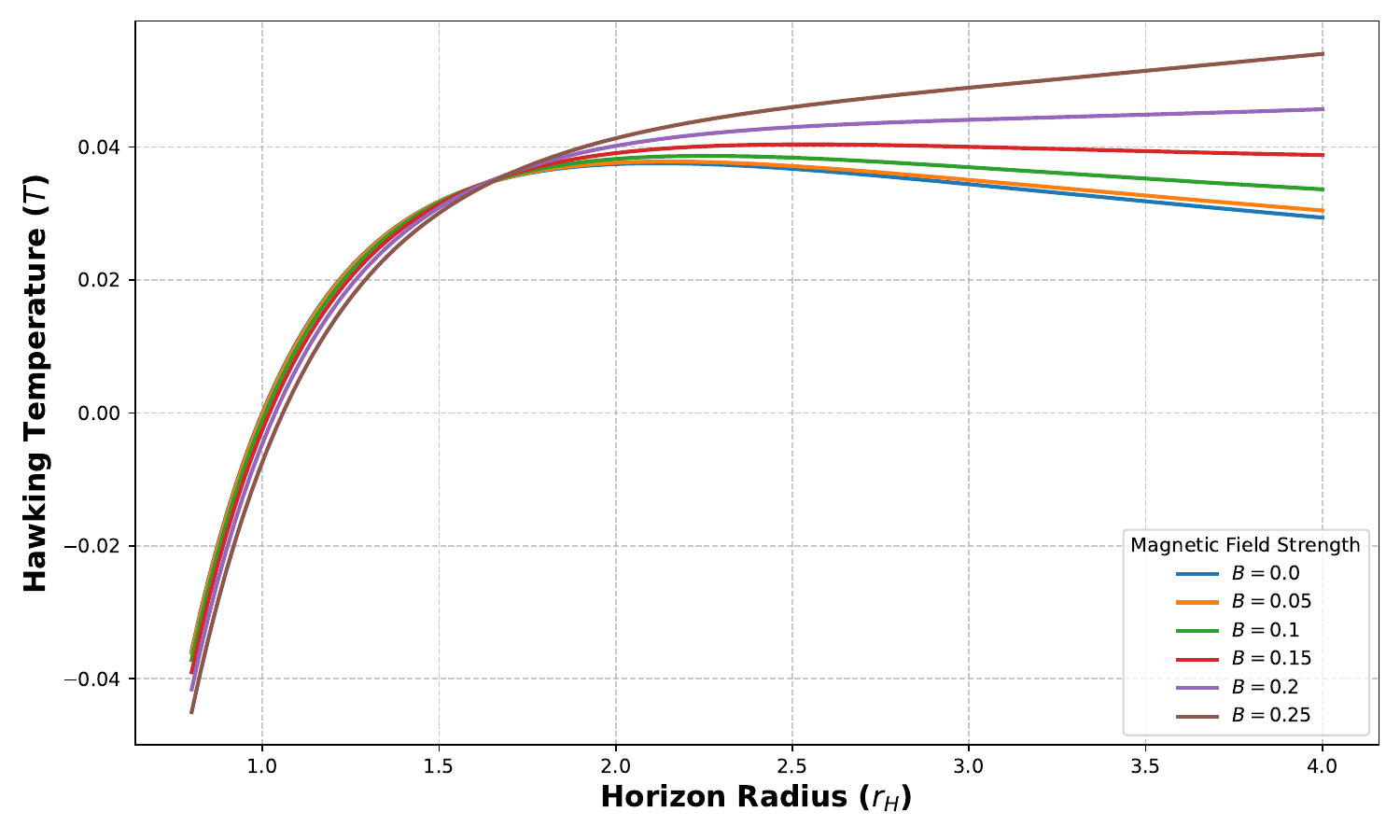}}
        \end{subfigure}
    \caption{Variation of Temperature ($S$) of KBR BH with (a) magnetic field ($B$) for different  values of rotation parameter ($\alpha$); (b) horizon radius ($r_H$) for different  values of rotation parameter ($\alpha$); (c) horizon radius ($r_H$) for different  values of magnetic field ($B$). Here, we consider $m=1$, $B=0.2$ and $\alpha=0.5$ where we kept those parameters as constant.}
    \label{fig:temperature}
\end{figure}
Fig.~\ref{fig:temperature} shows the graphical representation of the Hawking temperature $T$ against the magnetic field strength $B$ and the
horizon radius $r_H$.
Fig.~\ref{fig:temperature}(a) reveals the thermal response of the KBR spacetime across various spin parameters ($\alpha$). In contrast to the entropy profile, the temperature exhibits a continuous rise for all values of $\alpha$ as the magnetic field intensifies. This thermal behavior occurs because the surrounding magnetic energy density steepens the gravitational gradient right at the event horizon, directly increasing the BH's surface gravity ($\kappa$).
The Hawking temperature $T$ as a function of the horizon radius $r_H$ is shown in Figs.~\ref{fig:temperature}~(b) and (c). The resulting curves of the sub figure \ref{fig:temperature}~(b) exhibit the classic thermodynamic signature of evaporating BHs: a strict inverse relationship between size and temperature. As the horizon radius contracts, the gravitational gradient right at the boundary steepens violently, causing the Hawking temperature to diverge. This indicates that as the BH shrinks, it radiates thermal energy with extreme intensity. In contrast, as the horizon expands, the surface gravity weakens and the BH naturally cools. The stacked curves for the varying rotation parameters $\alpha$ provide a clear visualization of the cooling effect of the angular momentum. At any fixed horizon size, a higher spin parameter lowers the entire thermal curve. This confirms that, while the magnitude of the external magnetic background increases the overall temperature of the system, the BH's own intrinsic rotation acts as a powerful thermodynamic suppressant.\\
The plot in Fig.~\ref{fig:temperature}~(c) demonstrates the variation of the Hawking temperature $T$ against the arbitrary horizon radius $r_H$ for various magnetic field strengths $B$, with a constant rotation parameter $\alpha = 0.5$. With respect to the general shape of the curves at smaller radii, a clear inverse relationship can be observed: as the horizon radius becomes larger, the temperature drops. This represents the classic thermodynamic behavior of BHs, where smaller BHs radiate intensely and larger ones are much colder. The unique trend here is the thermal shift caused by the magnetic field. For any fixed horizon radius, a stronger external magnetic field results in a higher Hawking temperature. This happens because the surrounding magnetic background steepens the gravitational gradient (surface gravity) right at the event horizon, forcing the BH to radiate at a hotter thermal state than it would in an empty vacuum. Furthermore, as the horizon expands, a distinct thermodynamic shift occurs. After crossing a certain radius threshold, the temperature overcomes the standard geometric cooling and begins to increase indicates that the omnipresent magnetic energy density structurally dominates the spacetime at large scales, overpowering the standard temperature drop and driving the surface gravity back up.

\section{Result and discussion} \label{sec6}
In this study, we analyzed the gravitational deflection of the massless particle due to the rotating BH immersed in a uniform magnetic field (KBR BH) using the material medium approach. We also analyzed the thermodynamical behavior of this BH. Basically, we investigated the effect of a uniform magnetic field over the spin parameter of the BH, as rotation and background electromagnetic field change the structure of spacetime in meaningful ways. The remarkable results of this work are as follows:
\begin{itemize}
    \item The magnetic field pushed the event horizon outward over the Kerr BH, whereas the inner Cauchy horizons stay very close to each other, almost overlapping for slow rotation of the BH.
    \item The frame dragging effect shows the opposite behavior with respect to the Bh spin, while BHs without spin make no difference with the direction in which the photon moves. The spin of the BH increases the dragging effect for a prograde photon and decreases for a retrograde photon. But at higher magnetic field strengths, the overall frame-dragging  effect increases and decreases for prograde and retrograde motions, respectively, regardless of how fast the BH is spinning. At higher distances, the structural impact of the external magnetic field is observed. For an unmagnetized BH, the photon paths continuously decay toward zero, as the rotational drag fades in the empty space. But in case of higher magnetic field, the curves shift higher (for prograde) or lower (for retrograde) and flatten into permanently elevated or sunken, respectively. 
    \item As a prograde photon must navigate a significantly denser optical path, facing more optical resistance than a retrograde photon causes the refractive index to be higher than that of their corresponding retrograde photons. However, with the magnetic field, the refractive index decreases. This indicates that the magnetic field as a constant source of optical density causes the overall refractive index of the surrounding space, creating a denser optical baseline even before considering the rotational effect of the BH.
    \item Deflection curves also confirm the effect of denser optical path in its prograde motion compared to its retrograde motion. In the weak magnetic field regime, the magnetic field actively adds to the BH’s gravity, making the bending of light stronger.
    \item The curves without magnetic field (Kerr and Schwarzschild) behaves like a normal, flat vacuum at large distances, whereas the refractive indices of magnetized curves (SBR and KBR) keep dropping towards $n=0.0$ proving that the magnetic field ($B$) permanently changes the space far away, preventing it to act as a normal flat vacuum.
    \item The obtained results are also compared with those of the Kerr BH~$(B=0)$. The deflection angle for the KBR BH exceeds that of the Kerr spacetime in both the prograde and retrograde cases. This behavior indicates that the external magnetic field modifies the effective optical properties of the spacetime, resulting in enhanced gravitational lensing within the material medium framework.
    \item From thermodynamical investigation, it is clear that entropy monotonically decreases with magnetic field strength and rotation parameter, which confirms that the external magnetic environment and the internal angular momentum work together to geometrically restrict the overall information-storing capacity of the BH.  
    \item The entropy naturally increases with horizon radius  as a larger BH has a larger surface area, but the baseline entropy slightly increases with rotation parameter and significantly flattened with increasing external magnetic field.
    \item In contrast to entropy, the Hawking temperature increases with a uniform magnetic field but decreases with spin parameter because the magnetic energy density increases the BH’s surface gravity.
    With the expansion of the horizon, the surface gravity weakens and the BH naturally cools.
\end{itemize}

Thus, the magnetic background acts as an atypical optical medium, fundamentally altering the effective refractive index and distinguishing the gravitational lensing signatures of KBR BH from those of a standard BH.
\section*{Acknowledgments}
Authors, SR and AT express sincere and deep gratitude to the Department of Physics, NITA, for providing the necessary research environment to complete this work. The author, SK, sincerely acknowledges IMSc for providing exceptional research facilities and a conducive environment that facilitated his work as an Institute Postdoctoral Fellow. One of the authors, HN, would like to thank IUCAA, Pune, for the support under its associateship program, where a part of this work was done. The author H.N. acknowledges financial support from the Anusandhan National Research Foundation (ANRF), through the Science and Engineering Research Board (SERB) Core Research Grant (Grant No. CRG/2023/008980).
\section{Appendix A}
\begin{eqnarray*}
n(r) &=& \frac{r^2}{Q'(r)} \left[ 1 + 2\alpha  \left( \frac{r^2}{Q'(r)} - 1 \right)\frac{d\phi}{c dt} \right]^{-1/2}\nonumber\\
&=&  \frac{r^2}{Q'(r)} \left[ 1 + \epsilon \right]^{-1/2},
\end{eqnarray*}
where $\epsilon = 2\alpha \frac{d\phi}{c dt} \left( \frac{r^2}{Q'(r)} - 1 \right)$.
Since we consider the weak-field regime where $\alpha \ll r$, allows us to simplify the expression using the binomial approximation and with this approximation, the refractive index reduces to:

\begin{equation*}
n(r) \approx \frac{r^2}{Q'(r)} \left[ 1 - \alpha \left(\frac{r^2}{Q'(r) } - 1\right)\frac{d\phi}{cdt} \right]
\end{equation*}

Now, we express the metric function $Q^{\prime}(r)$, expressed in eqn. (\ref{eqn_Q_prime}) as
\begin{eqnarray*}
    Q^{\prime}(r)&=&( 1 + B^2 r^2 )(a r^2 - br)\\
    &=& a r^2 - b r + a B^2 r^4 - b B^2 r^3\\
    &\approx& a r^2 - b r + a B^2 r^4\\
    &=& a r^2 \left(1 - \frac{b}{a r} + B^2 r^2 \right).
\end{eqnarray*}
with $a=\left(1-\frac{B^2 I_2 m^2 }{I_1^2} \right)$ and $b=\frac{2 I_{2} m }{ I_{1} }$ and neglecting the sub leading term $-b B^2 r^3$.

Using the approximation $\frac{1}{1 + x} \approx 1 - x$, where $x = -\frac{b}{a r} + B^2 r^2$, we obtain:
\begin{equation*}
\frac{r^2}{Q'(r)} \approx \frac{1}{a}\left(1 + \frac{b}{a r} - B^2 r^2 \right).
\end{equation*}
Thus, the refractive index becomes:
\begin{widetext}
\begin{eqnarray*}
n(r) &=& \frac{1}{a}\left(1 + \frac{b}{a r} - B^2 r^2 \right) \left[ 1 - \alpha \left( \left(\frac{1}{a} - 1\right) + \frac{b}{a^2 r} - \frac{B^2 r^2}{a} \right)\frac{d\phi}{cdt} \right]\\
&=& \frac{1}{a} \Big[ 1 + \frac{b}{a r} - B^2 r^2 - \alpha Y(r)\frac{d\phi}{dt} - \left(\frac{b}{a r} - B^2 r^2\right)\alpha Y(r)\frac{d\phi}{cdt} \Big]\\
&=& \frac{1}{a} \left[ 1 + \frac{b}{a r} - B^2 r^2 - \alpha Y(r)\frac{d\phi}{cdt} \right],
\end{eqnarray*}    
\end{widetext}
where we have neglected the higher-order terms and defined $Y(r) = (\frac{1}{a} - 1) + \frac{b}{a^2 r} - \frac{B^2 r^2}{a}$.

\bibliography{mainKBR}
\bibliographystyle{ieeetr}
\end{document}